\documentclass[11pt]{article}
\pdfoutput=1
\usepackage{jheppub}

\usepackage{enumerate}
\usepackage{amsmath,amssymb,epsfig,cancel}
\usepackage{color}
\usepackage{subfig}
\usepackage{bbold}
\usepackage{graphicx}
\usepackage{physics}
\usepackage{caption}
\usepackage{braket}
\usepackage[english]{babel}

\newcommand{\eqa}{\overset{\mathrm{\alpha'=1}}{=\joinrel=}}

\newcommand{\case}[1]{%
  \sbox0{\bfseries Case #1 }%
  \noindent\hangindent=\wd0\box0\ignorespaces
}

\newcommand{\theorem}[1]{%
  \sbox0{\bfseries Theorem #1 }%
  \noindent\hangindent=\wd0\box0\ignorespaces
}

\newcommand{\step}[1]{%
  \sbox0{\bfseries Step #1 }%
  \noindent\hangindent=\wd0\box0\ignorespaces
}

\newcommand{\cc}[1]{%
  \sbox0{\bfseries ($c_{#1}$) }%
  \noindent\hangindent=\wd0\box0\ignorespaces
}

\newcommand{\cci}[1]{%
  \sbox0{\bfseries ($\infty_{#1}$) }%
  \noindent\hangindent=\wd0\box0\ignorespaces
}

\title{Non-integrability on $\text{AdS}_3$ supergravity backgrounds}

\author{Kostas Filippas,}

\affiliation{Department of Physics, Swansea University, Swansea SA2 8PP, United Kingdom}

\emailAdd{kphilippas@hotmail.com} 

\abstract{We investigate classical integrability on two recently discovered classes of backgrounds in massive IIA supergravity. These vacua are of the form AdS$_3\,\times\,$S$^2\times\mathbb{R}\,\times\,$CY$_2$, they preserve small $\mathcal{N}=(0,4)$ supersymmetry and are associated with D8$-$D6$-$D4$-$D2 Hanany-Witten brane set-ups. We choose an appropriate string embedding and use differential Galois theory on its associated Hamiltonian system, intending to produce the conditions under which Liouvillian solutions may occur. By constraining the parameters of the system according to the consistency of the associate brane set-ups we prove that no such conditions exist, yielding the complete non-integrability of these vacua. That is, up to the trivial cases where the background reduces to the Abelian and non-Abelian T-dual of AdS$_3\,\times\,$S$^3\times\,$T$^4$.
\\[5pt]
 }
 
\keywords{ Integrability, differential Galois theory, AdS$_3$, supergravity.\\[30pt]}

\subheader{}
\vspace{3cm}
\dedicated{Dedicated to the memory of Zackie Oh!.}

\begin{document}
\def\Tr{{\textrm{Tr}}}




\maketitle


\section{Introduction}
Integrability possesses an essential role in modern field theory. Not only it reveals a rich structure of conserved quantities that shape the physics of the system, but it also states that the theory is solvable for any choice of the coupling constant. Since holography relates the worldsheet theory of the superstring to a quantum field theory, integrable structures in string theory have won a prominent role in leading the way to new integrable gauge theories, \cite{Beisert:2010jr,Torrielli:2016ufi,Zarembo:2017muf}. Even the most successful calculations on the standard AdS/CFT correspondence, between AdS$_5\,\times$ S$^5$ supergravity and $\mathcal{N}=4$ super Yang-Mills theory, rely on the complete integrability of the system.\vspace{0.15cm}

However, spotting integrable structures can prove to be quite a challenging task. Integrability depends on the existence of a Lax connection on the cotangent bundle of the theory, while no standard recipe is provided to acquire such a construction. In fact, there is not even an a priori reason to decide whether such a connection does exist. That is, unless we acknowledge the theory to be non-integrable. Therefore, integrable systems are mainly obtained as structure-preserving deformations of known integrable theories, \cite{Lunin:2005jy,Sfetsos:2013wia,Delduc:2014kha,Borsato:2016pas}.\vspace{0.1cm}

Through the limitations of the classic methods of integrability, analytic non-integrability manifests itself in a dialectic way. Considering Hamiltonian systems of equations, analytic non-integrability makes use of Galois theory on differential equations to produce a statement on the structure of these systems. The arguments of differential Galois theory on second order, ordinary, linear differential equations were brought to an algebraic form by Kovacic \cite{Kovacic}, who also provided an explicit algorithm that produces the Liouvillian solutions of such equations, if any.\vspace{0.1cm}

In terms of supergravity, we choose a string embedding that produces the kind of differential equations of motion that can be examined under Kovacic's theorem, \cite{Zayas:2010fs,Basu:2011fw,Basu:2012ae,Chervonyi:2013eja,Stepanchuk:2012xi,Giataganas:2013dha,Giataganas:2014hma,Asano:2015eha,Ishii:2016rlk,Hashimoto:2016wme,Giataganas:2017guj,Roychowdhury:2017vdo,Banerjee:2018ifm,Roychowdhury:2019olt,Akutagawa:2019awh,Nunez:2018ags,Nunez:2018qcj,Filippas:2019puw}. Since an integrable theory has all of its dynamical sectors integrable, then every possible string configuration must echo integrable dynamics. Even a single sector exhibiting non-integrable behavior is enough to declare a supergravity vacuum as non-integrable. Therefore, we choose an embedding complicated enough to provoke the possibly non-integrable structure of the background but, at the same time, simple enough to produce the kind of differential equations we can examine under differential Galois theory.\vspace{0.15cm}

On another approach, S-matrix factorization on the worldsheet theory of the string was used to provide certain conditions of non-integrability, \cite{Wulff:2019tzh,Wulff:2017lxh,Wulff:2017vhv,Wulff:2017hzy}, while very recently a reconciliation began to arise between both non-integrability tools, \cite{Giataganas:2019xdj}.\vspace{0.1cm}

The present work, which employs differential Galois theory, comes as advertised and proves a recently discovered AdS$_3$ supergravity family, \cite{Lozano:2019emq,Lozano:2019jza,Lozano:2019zvg,Lozano:2019ywa}, to be classically non-integrable. That is, up to the trivial cases where the background reduces to the Abelian and non-Abelian T-dual of AdS$_3\,\times\,$S$^3\times\,$T$^4$. These massive IIA vacua are classified in \cite{Lozano:2019emq} in two distinct classes of backgrounds, from which we consider certain solutions of the form AdS$_3\,\times\,$S$^2\times\mathbb{R}\,\times\,$CY$_2$ as in \cite{Lozano:2019zvg}. The solutions preserve small $\mathcal{N}=(0,4)$ supersymmetry and are associated with D8$-$D6$-$D4$-$D2 Hanany-Witten brane set-ups. Holography suggests these backgrounds to be dual to two-dimensional quiver quantum field theories. Special holographic features of the AdS$_3$/CFT$_2$ duality over the solutions we consider were studied in \cite{Speziali:2019uzn}. Other warped massive IIA AdS$_3$ supergravities, associated with similar brane set-ups, were introduced in \cite{Dibitetto:2017klx,Dibitetto:2018iar}, while an extensive study of two-dimensional $\mathcal{N}=(0,4)$ quiver gauge theories was performed in \cite{Hanany:2018hlz}.\vspace{0.1cm}

At the same time, this article also aims to clarify the proper use of Kovacic's theorem on parametrized differential equations. In particular, we emphasize that failure of Kovacic's algorithm $-$ which is implemented in every algebra software $-$ on a parametrized equation does \textit{not} imply absence of Liouvillian solutions. It just states that \textit{not all} choices of the parameters lead to an integrable equation. It does certainly \textit{not} say that there are no particular selections among them that lead to integrability. Hence, if full generality on the parameters is demanded, then failure of Kovacic's algorithm indeed declares the non-integrability of the system. On the other hand, if the problem allows its parameters to be adjustable, no such statement can be made.\vspace{0.1cm}

In the latter case, we \textit{must} enforce the full power of Kovacic's theorem and go over its analytic algorithm by hand. If special parameter selections (that lead to an integrable structure) exist, then Kovacic's analytic algorithm will find them all, along with their associated solutions. If there are no such selections, then we can \textit{safely} declare our system as non-integrable.\vspace{0.1cm}

This is exactly what happens in our case. The AdS$_3$ supergravity family we consider is defined on general parameters whose adjustment equals picking different supergravity backgrounds. Therefore, the failure of Kovacic's algorithm here just states that not all possible backgrounds are integrable. It does not say that there are no integrable ones, among the whole family. But this is to be expected. It is the possible special combinations of these parameters, i.e. the particular supergravity backgrounds, that we are interested in. By demanding consistency on the supergravity brane set-ups, we show that the parameters are constrained in such a way that no integrable backgrounds of this supergravity family can exist. That is, as restated, up to the trivial cases where the background reduces to the Abelian and non-Abelian T-dual of AdS$_3\,\times\,$S$^3\times\,$T$^4$.\vspace{0.1cm}

The plan of this paper is as follows. In Section \ref{section1}, we present the backgrounds of the form AdS$_3\,\times\,$S$^2\times\mathbb{R}\,\times\,$CY$_2$ in a general manner and give a qualitative picture of their features. In Section \ref{section2}, we construct our string embedding and produce its equations of motion. By choosing a simple solution of these equations, we find the dynamical fluctuations around it. In Sections \ref{ClassIsection},\ref{ClassIIsection} we independently study the two classes of the AdS$_3$ backgrounds, by applying differential Galois theory on their associated fluctuations. Each class corresponds to a different kind of brane set-ups and, thus, exhibits different restrictions on its background parameters. By employing Kovacic's analytic algorithm, we show that in both supergravity classes these restrictions forbid integrability for all the possible backgrounds. Finally, in Section \ref{epilogue}, we summarize our results and give a review of our method as a concrete non-integrability tool.

\section{AdS$_3\,\times\,$S$^2\times\mathbb{R}\,\times\,$CY$_2$ \;\;vacua}\label{section1}
Let us outline the supergravity vacua that we are about to consider. It is essential to understand the basic aspects of these backgrounds, since it is the physical restrictions on their parameters that will ultimately decide the fate of their (non-) integrability.\vspace{0.1cm}

The massive IIA supergravity vacua first constructed in \cite{Lozano:2019emq} split in two distinct classes, Class I and II. From each class, we pick the solutions of the form AdS$_3\,\times\,$S$^2\times\mathbb{R}\,\times\,$CY$_2$ as in \cite{Lozano:2019zvg}. From now on, Class I and II will indicate this particular choice. Both classes have NS-NS sector, in string frame,
\begin{equation}
\begin{split}
\dd s^2\:=\:f_1\,\dd s^2_{\mbox{\tiny AdS$_3$}}+f_2\,\dd s^2_{\mbox{\tiny S$^2$}}+\frac{\dd\rho^2}{f_1}+f_3\,\dd s^2_{\mbox{\tiny CY$_2$}}\hspace{2cm}\\[10pt]
B_2\:=\:f_4\,\mbox{vol}_{\mbox{\scriptsize S$^2$}}\hspace{2cm}e^{-\Phi}\:=\:f_5\hspace{2cm}f_i\:=\:f_i\left(u,h_4,h_8\right)\label{GeneralAdS3background2}
\end{split}
\end{equation}\\
where $u,h_4,h_8$ are functions of the coordinates $\lbrace\rho,\mbox{CY}_2\rbrace$, left to be defined. The RR sector, consisting of $F_0$, $F_2$ and $F_4$, won't be needed here. These backgrounds enjoy a bosonic SL(2) $\times$ SU(2) isometry, they have eight supercharges and were proposed to be dual to $\mathcal{N}=(0,4)$ CFTs in two dimensions. Here we will consider the solutions on which the symmetries of CY$_2$ are globally respected. This restricts the internal Calabi-Yau manifold to be either

\begin{equation}
\mbox{CY}_2\:=\:\mbox{T}^4\hspace{1cm}\mbox{or}\hspace{1cm}\mbox{CY}_2\:=\:\mbox{K3}
\end{equation}\\
and the warp factors to be $f_i=f_i(\rho)$, i.e. $u=u(\rho)$, $h_4=h_4(\rho)$ and $h_8=h_8(\rho)$. The warp factor dependence on these functions will be specified for each supergravity class accordingly in the sections to follow. Preservation of the $\mathcal{N}=(0,4)$ supersymmetry and the Bianchi identities imply
\begin{equation}
u''(\rho)\:=\:0\hspace{2cm}h_4''(\rho)\:=\:h_8''(\rho)\:=\:0
\end{equation}\\
respectively. Therefore, all the defining functions are linear in $\rho$. In accordance with \cite{Lozano:2019emq}, we parametrize them as

\begin{equation}
u(\rho)\:=\:c_2+c_3\rho\hspace{1.3cm}h_4(\rho)\:=\:c_4+c_5\rho\hspace{1.3cm}h_8(\rho)\:=\:c_1+F_0\rho\label{uh4h8general}
\end{equation}\\
where all $c_i$ are real. For the new solutions to be associated with Hanany-Witten brane set-ups, these funtions are defined piecewise on the intervals $\rho\in[2\pi k,2\pi(k+1)]$, $k\in\mathbb{Z}$. Imposing that the functions vanish at $\rho=0$ where the space begins, we get

\begin{equation}
h_4(\rho)\:=\:\Upsilon\left\lbrace\begin{array}{cc}
\frac{c_5^0}{2\pi}\rho &\hspace{1cm}0\leq\rho\leq2\pi\\
c_4^k+\frac{c_5^k}{2\pi}(\rho-2\pi k) &\hspace{1cm}2\pi k\leq\rho\leq2\pi(k+1)\\
c_4^P+\frac{c_5^P}{2\pi}(\rho-2\pi P) &\hspace{1cm}2\pi P\leq\rho\leq2\pi(P+1)
\end{array}\right.\label{h4piecewise}
\end{equation}
\begin{equation}
h_8(\rho)\:=\:\left\lbrace\begin{array}{cc}
\frac{F_0^0}{2\pi}\rho &\hspace{1cm}0\leq\rho\leq2\pi\\
c_1^k+\frac{F_0^k}{2\pi}(\rho-2\pi k) &\hspace{1cm}2\pi k\leq\rho\leq2\pi(k+1)\\
c_1^P+\frac{F_0^P}{2\pi}(\rho-2\pi P) &\hspace{1cm}2\pi P\leq\rho\leq2\pi(P+1)
\end{array}\right.\label{h8piecewise}
\end{equation}\\
and $u(\rho)=\frac{c_3}{2\pi}\rho$. $\Upsilon$ is just a constant that may be normalized conveniently. The first derivatives of $h_4$, $h_8$ present discontinuities at $\rho=2k\pi$ where D4 and D8 branes are located\footnote{We omit to present the explicit dependence of the RR sector to $h_4$,$h_8$ (which, like the NS sector, differs for each class of vacua) to avoid unnecessary formulas. However, the restless reader is prompted to \cite{Lozano:2019emq} for details or to \cite{Lozano:2019zvg} for a clearer review.}, while $u''=0$ across all intervals as dictated by global supersymmetry. The discontinuities in the RR sector, that are interpreted as localized branes along $\rho$, modify the Bianchi identities appropriately with delta functions. Note that in order for supergravity to be trustable, $\lbrace c_1,..,c_5,F_0,P\rbrace$ have to be large.\\

Continuity of the NS-NS sector implies continuity of the $h_4$, $h_8$ functions across the $\rho$ intervals. This leads to

\begin{equation}
c_4^{k+1}\:=\:c_4^k+c_5^k\hspace{2cm}c_1^{k+1}\:=\:c_1^k+F_0^k
\end{equation}\\
which in turn gives

\begin{equation}
c_4^{k+1}\:=\:\sum_{j=0}^kc_5^k\hspace{2cm}c_1^{k+1}\:=\:\sum_{j=0}^kF_0^k
\end{equation}\\
In order to gain a better feel on the parameters $\lbrace c_1,..,c_5,F_0\rbrace$ we consider, as an example, the RR charges of Class I supergravity, in the intervals $[2\pi k,2\pi(k+1)]$. For $\alpha'=g_s=1$, a D$p$-brane is charged under $Q_{Dp}=(2\pi)^{p-7}\int_{\Sigma_{8-p}}F_{8-p}$, thus in our set-up they read\footnote{$F_0^k$ is $F_0$ in the $k$-th interval. Whenever we loose the $k$ subscript we will mean $F_0^k$.}

\begin{equation}
\begin{split}
Q_{D8}\:&=\:F_0^k\hspace{4.4cm}Q_{D6}\:=\:\frac{1}{2\pi}\int_{\mbox{\scriptsize S$^2$}}F_2\sim \;c_1^k\\
Q_{D4}&=\frac{1}{8\pi^3}\int_{\mbox{\scriptsize CY$_2$}}F_4\sim \;c_5^k\hspace{2cm}Q_{D2}\:=\:\frac{1}{32\pi^5}\int_{\mbox{\scriptsize CY$_2\times$S$^2$}}F_6\sim \;c_4^k
\end{split}
\end{equation}\\
and $Q_{NS}=\frac{1}{4\pi^2}\int_{\rho\times\mbox{\scriptsize S}^2}H_3=P+1$. A study of the Bianchi identities reveals that no explicit D2 and D6 branes are present in the geometry, just their fluxes\footnote{This is true when the worldvolume gauge field on the D8, D4 branes is absent. When it is on, there is D6 and D2 flavor charge induced on the D8's and D4's. See the appendix B of \cite{Lozano:2019zvg} for details.}. This associates their amount, $c_4^k$ and $c_1^k$ respectively, with the ranks of the (color) gauge groups in the dual field theory. On the other hand, as restated, D8 and D4 branes do exist in the geometry and modify the Bianchi identities by a delta function. Thus, $F_0^k$ and $c_5^k$ are associated with the ranks\footnote{The rank is a positive number. If the slope is negative, that is related to the orientation of the branes.} of the (flavor) global symmetries of the dual field theory.\\

\begin{figure}[h!]
    \centering
    {{\includegraphics[width=9cm]{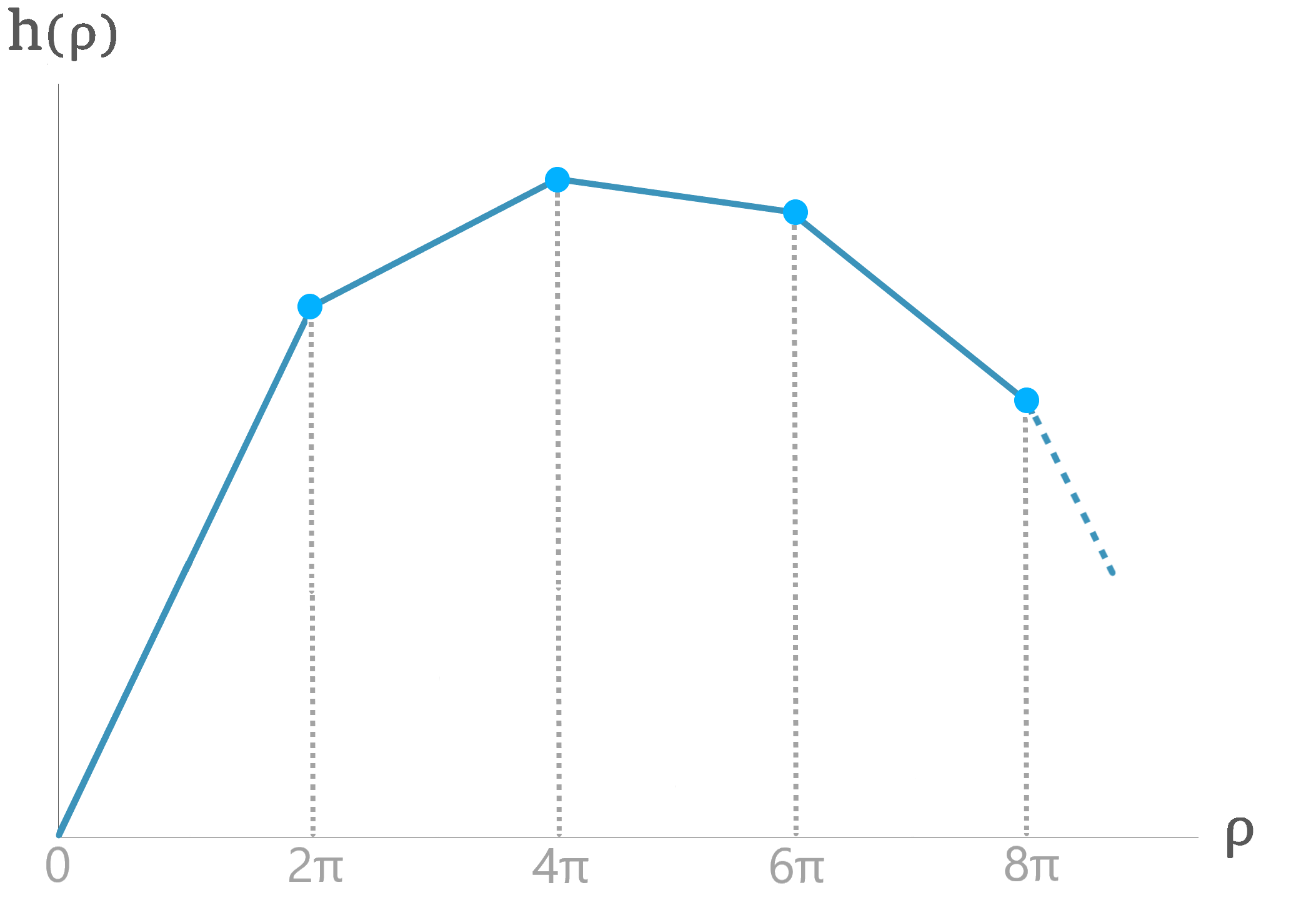} }}%
\caption{An example of a linear function $h_{4,8}(\rho)$. This kind of function is defined piecewise on every interval $\rho\in[2\pi k,2(k+1)\pi]$, while it decreases in slope along the $\rho$ dimension.}
\label{figure1}
\end{figure}

Realizing the $h_4$ and $h_8$ pieces across the $\rho$ dimension as blocks of gauge and flavor groups in the dual two-dimensional quantum field theory, we assembly them to quiver gauge theories. Then, cancellation of their gauge anomalies implies

\begin{equation}
N_{D8}^{[k-1,k]}\:=\:F_0^{k-1}-F_0^k\hspace{2cm}N_{D4}^{[k-1,k]}\:=\:c_5^{k-1}-c_5^k
\end{equation}\\
For the $h_4$, $h_8$ functions this translates to decreasing slopes\footnote{Or slopes that remain the same across intervals, giving no flavor branes between them.}, $c_5^k$ and $F_0^k$ respectively, as $\rho$ increases. Thus, any of these functions draws a piecewise linear curve of decreasing slope, as in Figure \ref{figure1}.

While the present section provides a consistent summary of this particular AdS$_3$ supergravity and its dual quiver field theory, the reader is prompted to \cite{Lozano:2019emq} for details on the construction of the solutions, to \cite{Lozano:2019jza} for an overview and to \cite{Lozano:2019zvg} for a deeper dive into the quiver realization.\\

\section{String dynamics on AdS$_3\,\times\,$S$^2\times\mathbb{R}$}\label{section2}
The bosonic string dynamics is reflected on the non-linear $\sigma$-model, in conformal gauge,

\begin{equation}
S_P\:=\:\frac{1}{4\pi\alpha'}\int_\Sigma\dd^2\sigma\,\partial_aX^\mu\partial_bX^\nu\left(g_{\mu\nu}\eta^{ab}+B_{\mu\nu}\epsilon^{ab}\right)\label{PolyakovAction}
\end{equation}\\
where the string coordinates $X^\mu(\tau,\sigma)$ equation of motion is supplemented by the Virasoro constraint $T_{ab}=0$, where the worldsheet energy-momentum tensor is given by

\begin{equation}
T_{ab}\:=\:\frac{1}{\alpha'}\left(\partial_aX^\mu\partial_bX^\nu g_{\mu\nu}-\frac{1}{2}\eta_{ab}\eta^{cd}\partial_cX^\mu\partial_dX^\nu g_{\mu\nu}\right)\label{EMtensor}
\end{equation}\\
We desire a string embedding that produces ordinary differential equations as its equations of motion, so that we can apply differential Galois theory. This means that the string coordinates must be $X^\mu=X^\mu(\tau)$ or $X^\mu=X^\mu(\sigma)$, where $\tau, \sigma$ are the worldsheet coordinates. Since the search of (non-) integrability requires bringing dynamics to the test, we like our soliton to have as much stringy character as possible, according always to the above restriction $X^\mu=X^\mu(\sigma)$. Thus, we wrap it around all cyclic coordinates available.\\

Both Class I and II of the AdS$_3$ supergravity we consider consist of the NS-NS sector, in the string frame,

\begin{equation}
\begin{split}
\dd s^2\:&=\:f_1\,\dd s^2_{\mbox{\tiny AdS$_3$}}+f_2\,\dd s^2_{\mbox{\tiny S$^2$}}+\frac{\dd\rho^2}{f_1}+f_3\,\dd s^2_{\mbox{\tiny CY$_2$}}\\[10pt]
B_2\:&=\:f_4\,\mbox{vol}_{\mbox{\scriptsize S$^2$}}\hspace{1.5cm}e^{-\Phi}\:=\:f_5\label{GeneralAdS3background}
\end{split}
\end{equation}\\
where $f_i=f_i(\rho)$ are the various warp factors, left undefined for each supergravity class to be separately examined, and vol$_{\mbox{\tiny S$^2$}}=\sin\chi\,\dd\chi\wedge\dd\xi$. If global AdS$_3$ and S$^2$ with unit radii are expressed as

\begin{equation}
\begin{split}
\dd s^2_{\mbox{\tiny AdS$_3$}}\:&=\:-\cosh^2r\,\dd t^2+\dd r^2+\sinh^2r\,\dd\phi^2\\[10pt]
\dd s^2_{\mbox{\tiny S$^2$}}\:&=\:\dd\chi^2+\sin^2\chi\,\dd\xi^2
\end{split}
\end{equation}\\
then we set up our string embedding to be

\begin{equation}
\begin{split}
t=t(\tau)\hspace{0.7cm}r=r(\tau)\hspace{0.7cm}\phi=\nu\sigma\\[10pt]
\rho=\rho(\tau)\hspace{0.7cm}\chi=\chi(\tau)\hspace{0.7cm}\xi=\kappa\sigma
\end{split}\label{stringSoliton}
\end{equation}\\
where we wrapped the string $\nu$ and $\kappa$ times around the $\phi$ coordinate and the $\xi$ dimension, respectively. CY$_2$ dynamics was left out of the game, since it won't be eventually needed in the hunt of non-integrability. Note that it is the wrapping that provides the stringy, non-trivial behavior to the configuration. Without it we would just have point particle dynamics. Indeed, one of these winding modes will play a crucial role later on when we enforce differential Galois theory.\\

\subsection{Equations of motion}
Instead of the action (\ref{PolyakovAction}), it is more convenient working with its associated Langrangian density
\begin{equation}
\mathcal{L}\:=\:f_1\left(\cosh^2r\,\dot{t}^2-\dot{r}^2+\nu^2\sinh^2r\right)-\frac{\dot{\rho}^2}{f_1}-f_2\left(\dot{\chi}^2-\kappa^2\sin^2\chi\right)+2\kappa f_4\sin\chi\dot{\chi}
\end{equation}\\
where the dot implies derivation w.r.t the worldsheet time $\tau$. For our particular string embedding, the equations of motion for this Lagrangian are equivalent to those of the $\sigma$-model and read

\begin{equation}
\begin{split}
\dot{t}\:&=\:\frac{E}{f_1\,\cosh^2r}\\[10pt]
\ddot{r}\:&=\:-\frac{\nu^2f_1^2\,\sinh2r+2E^2\tanh r\,\sech^2r+2f_1\,f_1'\,\dot{r}\dot{\rho}}{2f_1^2}\\[10pt]
\ddot{\chi}\:&=\:-\kappa^2\cos\chi\,\sin\chi+\frac{\dot{\rho}\left(-f_2'\,\dot{\chi}+\kappa f_4'\,\sin\chi\right)}{f_2}\\[10pt]
\ddot{\rho}\:&=\:\frac{f_1'\left(-E^2\sech^2r+f_1^2(-\nu^2\sinh^2r+\dot{\rho}^2)\right)+f_1^2\left((-\kappa^2\sin^2\chi+\dot{\chi}^2)f_2'-2\kappa f_4'\sin\chi\dot{\chi}\right)}{2f_1}
\end{split}\label{EOM}
\end{equation}\\
where the dash on $f_i$'s implies derivation w.r.t their argument $\rho$. Notice that we have replaced the equation of motion for $t$ into the rest of the equations. These equations of motion are constrained by the worldsheet equation of motion, i.e. the Virasoro constraint

\begin{equation}
\begin{split}
2\,T_{\tau\tau}\:&=\:2\,T_{\sigma\sigma}\:=\:f_1\,\left(-\cosh^2r\,\dot{t}^2+\dot{r}^2+\nu^2\sinh^2r\right)+\frac{\dot{\rho}^2}{f_1}+f_2\,\left(\dot{\chi}^2+\kappa^2\sin^2\chi\right)\:=\:0\\
T_{\sigma\tau}\:&=\:0
\end{split}\label{VirasoroConstraints}
\end{equation}\\
This constraint holds regardless of the equations of motion and, thus, it is a primary constraint. The energy-momentum tensor is preserved on shell, $\nabla_aT^{ab}=0$, since $\partial_\tau T_{\tau\tau}=\partial_\sigma T_{\sigma\sigma}=0$ on the equations of motion (\ref{EOM}). Note, also, that the compliance of the worldsheet constraints with the equations of motion yield the consistency of our embedding.\\

In order to deeply appreciate our method and get a better grip on its physics, we break on through to the Hamiltonian formulation, by defining the conjugate momenta

\begin{equation}
p_t\:=\:2f_1\cosh^2r\,\dot{t}\hspace{1cm}p_r\:=\:-2f_1\dot{r}\hspace{1cm}p_\chi\:=\:-2f_2\dot{\chi}+2\kappa f_4\sin\chi\hspace{1cm}p_\rho\:=\:-\frac{2\dot{\rho}}{f_1}
\end{equation}\\
and the Hamiltonian density

\begin{equation}
\mathcal{H}\:=\:\frac{p_t^2}{4f_1\cosh^2r}-\frac{p_r^2}{4f_1}-\frac{p_\rho^2}{4(f_1)^{-1}}-\frac{\left(p_\chi-2\kappa f_4\sin\chi\right)^2}{4f_2}-\kappa^2f_2\sin^2\chi-\nu^2f_1\sinh^2r
\end{equation}\\
In this language, the Virasoro constraint is $\mathcal{H}=0$. Hamilton's equations on $\mathcal{H}$ and $p_i$ coincide, of course, with the Euler-Lagrange equations of motion (\ref{EOM}). Therefore, our string dynamics problem reduces to that of a particle in a non-trivial potential. In particular, the effective mass is defined by geometry through the kinetic terms, while the winding modes in the string perspective are realized as a potential on the particle.\\

\subsection{Normal Variational Equation}
While a system of involved differential equations of motion is unattractive to solve, there are always a few delicate ways to handle it. One of them is to look for a simple solution and expand around it, evaluating this way the dynamical behavior of the system. Stated otherwise, we look in the equations of motion for the simplest solution available by one of the variables and, given this solution, we study the fluctuations of the rest of the variables around it. We call such a fluctuation a Normal Variational Equation (NVE). \\

Taking up the equations of motion (\ref{EOM}), we easily see that their jet bundle prefers the point
\begin{equation}
r\:=\:\dot{r}\:=\:\ddot{r}\:=\:\chi\:=\:\dot{\chi}\:=\:\ddot{\chi}\:=\:0\label{TheChoice}
\end{equation}\\
which satisfies the $r$ and $\chi$ equations, while the one for $\rho$ becomes

\begin{equation}
\ddot{\rho}\:=\:\frac{f_1'}{2f_1}\left(\dot{\rho}^2-E^2\right)
\end{equation}\\
yielding the simple solution

\begin{equation}
\rho_{sol}=E\tau\label{rhosimplesol}
\end{equation}\\
where we omit an integration constant without loss of generality. Notice that vanishing all variables but $\rho$ is the simplest way to go, the rest of the choices leading to complicated solutions for $r$ or $\chi$.\\

Since the Virasoro constraint (\ref{VirasoroConstraints}) is essentially the equation of motion for the worldsheet metric and as such holds independently from the string coordinates' equations of motion, (\ref{EOM}), it should reflect the same physics, at least classically, if not a more constrained one. Indeed, enforcing the choice (\ref{TheChoice}) onto the Virasoro constraint we acquire
\begin{equation}
\dot{\rho}^2\:=\:E^2
\end{equation}\\
i.e. the same solution as (\ref{rhosimplesol}). Depending on the particular quality of a system, one can choose to seek for a simple solution on either the standard string equations of motion or on the Virasoro constraint. Regardless, any invariant plane we choose to fluctuate on must be a solution of both the string coordinates' equation of motion and the Virasoro constraint, in order for it to be consistent with our string embedding.\vspace{0.2cm}

Now, since the simple solution $\rho_{sol}$ is localized on the point (\ref{TheChoice}), then it is that point around which we study the fluctuations of $r,\chi$. Letting $r(\tau)=0+\epsilon\varrho(\tau)$ into the $r$ equation of motion in (\ref{EOM}), we expand for $\epsilon\rightarrow0$ and obtain its NVE at leading order as

\begin{equation}
\begin{split}
\ddot{\varrho}(\tau)+\mathcal{B}_\varrho(\tau)\dot{\varrho}(\tau)+\mathcal{A}_\varrho(\tau)\varrho(\tau)\:=\:0\hspace{1.7cm}\\[15pt]
\mathcal{B}_\varrho(\tau)\:=\:\frac{Ef_1'}{f_1}\bigg|_{\rho_{sol}}\hspace{1.5cm}\mathcal{A}_\varrho(\tau)\:=\:\frac{E^2+\nu^2f_1^2}{f_1^2}\bigg|_{\rho_{sol}}
\end{split}\label{NVEr}
\end{equation}\\
In the same manner, letting $\chi(\tau)=0+\epsilon x(\tau)$ we obtain the NVE for $\chi$ as

\begin{equation}
\begin{split}
\ddot{x}(\tau)+\mathcal{B}_x(\tau)\dot{x}(\tau)+\mathcal{A}_x(\tau)x(\tau)\:=\:0\hspace{1.7cm}\\[15pt]
\mathcal{B}_x(\tau)\:=\:\frac{Ef_2'}{f_2}\bigg|_{\rho_{sol}}\hspace{1.5cm}\mathcal{A}_x(\tau)\:=\:\frac{\kappa^2f_2-\kappa Ef_4'}{f_2}\bigg|_{\rho_{sol}}
\end{split}\label{NVEx}
\end{equation}\\
Using the change of variable $y=e^{\frac{1}{2}\int\mathcal{B}}z$ in the above differential equations, we deduce two new ones of the kind

\begin{equation}
z''\:=\:\mathcal{V}\,z\hspace{2cm}\mathcal{V}=\frac{1}{4}\left(2\mathcal{B}'+\mathcal{B}^2-4\mathcal{A}\right)\label{transformedNVE}
\end{equation}\\
where $y$ is Liouvillian if and only if $z$ is Liouvillian and, thus, no generality is lost. In this new variable, the NVEs for $r$ and $\chi$ read

\begin{equation}
\ddot{\varrho}\:=\:\mathcal{V}_\varrho\,\varrho\hspace{2.1cm}\mathcal{V}_\varrho\:=\:-\nu^2-\frac{E^2\left(4+(f_1')^2-2f_1f_1''\right)}{4f_1^2}\hspace{1.8cm}\label{newNVEr}
\end{equation}\\
\begin{equation}
\ddot{x}\:=\:\mathcal{V}_x\,x\hspace{2cm}\mathcal{V}_x\:=\:-\kappa^2-\frac{E\left(E(f_2')^2-2f_2(2\kappa(f_4')^2+Ef_2'')\right)}{4f_2^2}\label{newNVEx}
\end{equation}\\

Therefore, we end up with two linear, second order, ordinary differential equations. After defining $f_i(\rho_{sol})$ in every supergravity class, each $\mathcal{V}$ $-$ which we call the \textit{potential} $-$ will turn out to be a rational function of $\tau$. Hence, eventually, equations (\ref{newNVEr})-(\ref{newNVEx}) for $r$ and $\chi$ are of the appropriate form to be examined by differential Galois theory for Liouvillian integrability.\vspace{0.2cm}

Differential Galois theory on differential equations boils down to Kovacic's algorithm, \cite{Kovacic}. Kovacic provided three criteria on the pole structure of differential equations of the form (\ref{NVEr}) and (\ref{NVEx}) that decide if a Liouvillian solution can exist. These conditions are necessary but not sufficient for integrability. In other words, if none of these criteria is satisfied then we deduce with certainty that no Liouvillian solution exists. In that case, the dynamical sector under examination and, thus, the whole theory are non-integrable. On the other hand, even if one of the criteria is satisfied, then such a solution may exist and if it does then Kovacic's algorithm will find it. If the algorithm fails, no Liouvillian solution exists. A detailed analysis is found in Appendix \ref{appendixA}.\vspace{0.2cm}

In what follows, we employ the analysis of the present section to examine separately each AdS$_3\,\times\,$S$^2\times\mathbb{R}\,\times\,$CY$_2$ supergravity class of the form (\ref{GeneralAdS3background}), first presented in \cite{Lozano:2019emq}. After defining each class through the functions $f_i(\rho)$ and, thus, specify the corresponding NVEs, we intend to put Kovacic's theorem to the test.

\section{Class I backgrounds}\label{ClassIsection}
Given the general form of the NS-NS sector of the AdS$_3\,\times\,$S$^2\times\mathbb{R}\,\times\,$CY$_2$ massive IIA supergravity, at string frame, as
\begin{equation}
\begin{split}
\dd s^2\:=\:f_1\,\dd s^2_{\mbox{\tiny AdS$_3$}}+f_2\,\dd s^2_{\mbox{\tiny S$^2$}}+\frac{\dd\rho^2}{f_1}+f_3\,\dd s^2_{\mbox{\tiny CY$_2$}}\hspace{2cm}\\[10pt]
B_2\:=\:f_4\,\mbox{vol}_{\mbox{\scriptsize S$^2$}}\hspace{2cm}e^{-\Phi}\:=\:f_5\hspace{2cm}f_i\:=\:f_i\left(u,h_4,h_8\right)\label{GeneralAdS3background3}
\end{split}
\end{equation}\\
then the first supergravity class is defined by the warp factors

\begin{equation}
\begin{split}
f_1\:=\:\frac{u}{\sqrt{h_4h_8}}\hspace{1.5cm}f_2\:=\:f_1\frac{h_4h_8}{4h_4h_8+(u')^2}\hspace{1.5cm}f_3\:=\:\sqrt{\frac{h_4}{h_8}}\hspace{1.3cm}\\[10pt]
f_4\:=\:\frac{1}{2}\left(-\rho+\frac{uu'}{4h_4h_8+(u')^2}\right)\hspace{2cm}f_5\:=\:\frac{h_8^{\frac{3}{4}}}{2h_4^{\frac{1}{4}}\sqrt{u}}\sqrt{4h_4h_8+(u')^2}
\end{split}\label{ClassIwarps}
\end{equation}\\
For simplicity, we treat the functions $h$, $u$ in a general manner, as in (\ref{uh4h8general}), i.e.

\begin{equation}
u(\rho)\:=\:c_3\rho\hspace{1.3cm}h_4(\rho)\:=\:c_4+c_5\rho\hspace{1.3cm}h_8(\rho)\:=\:c_1+F_0\rho\label{uhClassI}
\end{equation}\\
since their piecewise character, (\ref{h4piecewise})-(\ref{h8piecewise}), can be always assumed. Meaning, whatever result we reach can be assumed to hold for any interval of these functions along the $\rho$ dimension.\vspace{0.2cm}

\begin{figure}[h!]
    \centering
    {{\includegraphics[width=9cm]{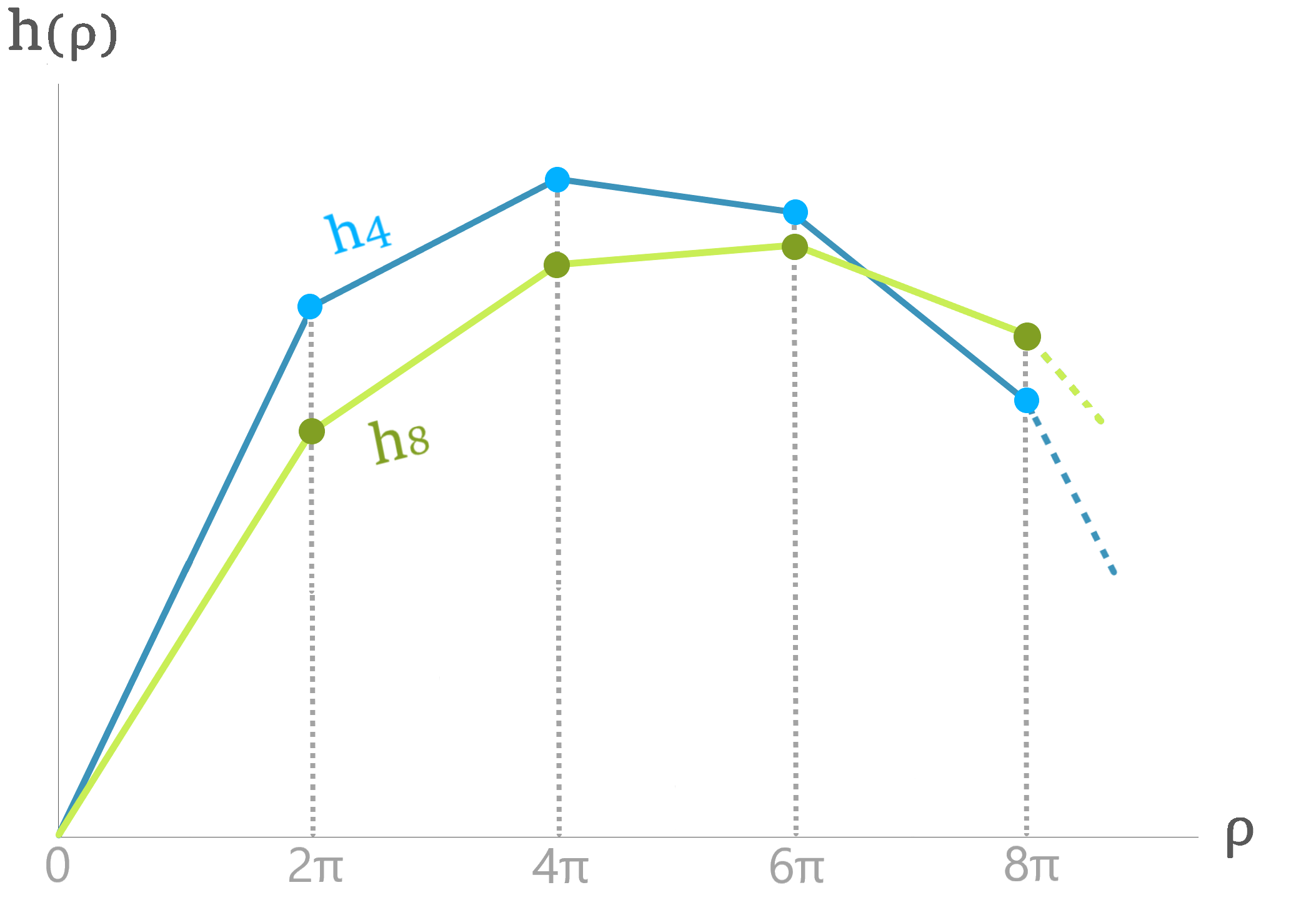} }}%
\caption{An example of the linear functions $h_{4,8}(\rho)$ in Class I backgrounds. These piecewise functions decrease in slope along $\rho$ and at least one of them (or both) has to vanish at the end of the dimension, $\rho_f$.}
\label{figure2}
\end{figure}

Notice that $h_4$ and $h_8$ can only vanish at the beginning and at the end of the $\rho$ coordinate. Otherwise, the background would degenerate and blow up at points along $\rho$. In fact, both of these functions vanish at $\rho=0$ and at least one of them has to vanish on the end of the $\rho$ dimension, $\rho_f$, for the space to end in a smooth fashion. Hence, $h_4$ and $h_8$ preserve their sign: they begin as positive piecewise linear curves and they remain this way, while their slope decreases along $\rho$. An example is drawn in Figure \ref{figure2}.\\

\subsection{Abelian T-dual of AdS$_3\,\times\,$S$^3\times\,$T$^4$}\label{sectionATDclassI}
Although we chose the functions $u$, $h_4$, $h_8$ such that Class I backgrounds begin and end in a smooth fashion, i.e. (\ref{uhClassI}) and Figure \ref{figure2}, it is worth breaking that rule for a brief moment. That is, we can trivially choose their most general form (\ref{uh4h8general}) to reduce to constant functions, i.e. $u=c_2$, $h_4=c_4$ and $h_8=c_1$. Then the background reduces to

\begin{equation}
\begin{split}
\dd s^2\:&=\:R^2\left(\dd s^2_{\mbox{\tiny AdS$_3$}}+\frac{1}{4}\,\dd s^2_{\mbox{\tiny S$^2$}}\right)+\frac{\dd\rho^2}{R^2}+\sqrt{\frac{c_4}{c_1}}\,\dd s^2_{\mbox{\tiny CY$_2$}}\hspace{2cm}\\[10pt]
B_2\:&=\:-\frac{\rho}{2}\,\mbox{vol}_{\mbox{\scriptsize S$^2$}}\hspace{2cm}\Phi\sim const.\label{NATDclassI}
\end{split}
\end{equation}\\
which is the Abelian T-dual (ATD) of AdS$_3\,\times\,$S$^3\times\,$T$^4$. The latter symmetric background is classically integrable, \cite{Babichenko:2009dk}. Hence, its Abelian T-dual, this duality being a canonical transformation, will preserve bosonic integrability. This last statement was formally elaborated in \cite{Borsato:2016pas}. Thus, the trivial choice of constant functions $u$, $h_4$, $h_8$, which is slightly outside the smooth choices we consider, leads to an integrable background.\\

Notice that we only picked CY$_2=$ T$^4$, since global metrics on K3 are not explicitly known. They should exist from Yau's theorem, but this fact is obviously useless w.r.t examining string dynamics on these surfaces. The same holds, of course, on the next subsection.

\subsection{Non Abelian T-dual of AdS$_3\,\times\,$S$^3\times\,$T$^4$}\label{sectionNATDclassI}
Before the general treatment, a provoking choice of parameters in (\ref{uhClassI}) is $c_1=c_4=0$, since then AdS$_3$ unwarps from the rest of the space and the background reduces to

\begin{equation}
\begin{split}
\dd s^2\:&=\:R^2\,\dd s^2_{\mbox{\tiny AdS$_3$}}+\left(\frac{R^2\rho^2}{R^4+4\rho^2}\right)\,\dd s^2_{\mbox{\tiny S$^2$}}+\frac{\dd\rho^2}{R^2}+\sqrt{\frac{c_5}{F_0}}\,\dd s^2_{\mbox{\tiny CY$_2$}}\hspace{2cm}\\[10pt]
B_2\:&=\:\left(-\frac{2\rho^3}{R^4+4\rho^2}\right)\,\mbox{vol}_{\mbox{\scriptsize S$^2$}}\hspace{2cm}\Phi\sim-\ln(1+\rho^2)\label{NATDclassI}
\end{split}
\end{equation}\\
where $R^2=\frac{c_3}{\sqrt{c_5F_0}}$. This particular background is the non Abelian T-dual (NATD) of AdS$_3\,\times\,$S$^3\times\,$T$^4$, having dualised one of the SU(2) subgroups of S$^3$, \cite{Sfetsos:2010uq}. The latter symmetric background is classically integrable, \cite{Babichenko:2009dk}. Hence, its non Abelian T-dual, this duality being a canonical transformation, will preserve bosonic integrability. Therefore, $c_1=c_4=0$ leads to an integrable background, (\ref{NATDclassI}), or, more generally, to an integrable interval of this class of backgrounds\footnote{Letting $c_1=c_4=0$ be true for all intervals, we inherit an overall NATD integrable theory. Letting it be true for a specific $\rho$-interval means that the background on this particular interval is an integrable NATD of AdS$_3\,\times\,$S$^3\times\,$T$^4$. Henceforth, we study all other cases except the trivial one where $c_1=c_4=0$ everywhere.}.\vspace{0.2cm}

Since this particular choice of parameters gives an integrable structure, this should be reflected on the corresponding $r$ and $\chi$ NVEs. Indeed, this is the case and the details are given in Appendix \ref{appendixNATD}.\vspace{0.2cm}

Recalling that $h_4$ and $h_8$ are defined piecewise in $\rho$, (\ref{h4piecewise})-(\ref{h8piecewise}), we realize that the choice $c_1=c_4=0$ reflects only the first interval, $[0,2\pi]$, of both the functions. That would be the first interval for both curves in Figure \ref{figure2}. Thus, we conclude that all possible geometries in this supergravity class begin as NATDs of AdS$_3\,\times\,$S$^3\times\,$T$^4$ with radius $R^2=\frac{c_3}{\sqrt{c_5F_0}}$ and are integrable in that part of their space.\vspace{0.2cm}

Then $h_4$ and $h_8$ drive along $\rho$ as positive functions of decreasing slope and, depending on the particular selection of their parameters $\lbrace c_i,F_0\rbrace$, they may give various backgrounds associated with appropriate brane set-ups. The positiveness of $\;h_4(\rho)=c_4+c_5\rho\;$ and $h_8(\rho)=c_1+F_0\rho\;$ combined with the decreasing slopes along $\rho$ mean that $c_1$ and $c_4$ are always non-negative,

\begin{equation}
c_1, c_4\geq0\hspace{2cm}c_1c_4\geq0\label{c1c4>0}
\end{equation}\\
while increasing (or staying the same) across the intervals\footnote{In case of confusion, $c_4$ and $F_0$ here represent the constants of $h_4$ and $h_8$ in a random interval. According to the piecewise definition (\ref{h4piecewise})-(\ref{h8piecewise}), these would reflect to the constants $c_4^k-c_5^kk$ and $c_1^k-F_0^kk$, respectively.}. This is equivalent to saying that each linear curve on every interval of Figure \ref{figure2} has a non-negative projection on the $\rho=0$ axis. Apart from providing a clearer picture on the overall brane set-up, this statement will define the outcome of the next section where we investigate integrability.\\

Expanding near $\rho\rightarrow0^+$ the space becomes AdS$_3\times\mathbb{R}^3\,\times\,$T$^4$, which is symmetric and integrable, as expected for any vicinity of an integrable background like (\ref{NATDclassI}). Hence, our study of (non-) integrability narrows down to all other intervals except that first NATD one and, from now on, it is those intervals that our study implies.\\

\subsection{NVE for $r$}
Let us begin our integrability analysis on the intervals next to the first NATD one, by first studying the string dynamics along $r$. Letting the warp factors (\ref{ClassIwarps}) roll on the NVE for $r$, (\ref{newNVEr}), we obtain

\begin{equation}
\ddot{\varrho}\:=\:\frac{Q_{\mbox{\tiny I}}}{\tau^2(\tau+\frac{c_4}{c_5E})^2(\tau+\frac{c_1}{F_0E})^2}\,\varrho\label{explicitNVEr}
\end{equation}\\
where $Q_{\mbox{\tiny I}}=Q_{\mbox{\tiny I}}(\tau^6,c_i,F_0,E)$ is a long polynomial in the numerator whose explicit form will not concern us. Now, the object that essentially needs to fall under our microscope is the potential $\mathcal{V}_\varrho$. Here, it comes with three poles of order two, $\lbrace\tau_1=0,\tau_2=-\frac{c_4}{c_5E},\tau_3=-\frac{c_1}{F_0E}\rbrace$ and it expands around $\tau\rightarrow\infty$ as

\begin{equation}
\mathcal{V}_\varrho^\infty\:=\:-(\frac{c_5F_0E^2}{c_3^2}+\nu^2)-\frac{(c_1c_5+c_4F_0)E}{c_3\tau}+\mathcal{O}\left(\frac{1}{\tau^2}\right)\label{Vrinfinity}
\end{equation}\\
exhibiting zero order behavior there. Thus, $\mathcal{V}_\varrho$ satisfies the first and second Kovacic's criteria, implying that the NVE (\ref{explicitNVEr}) may have Liovillian solutions. However, Kovacic's algorithm fails to solve it as it is.\\

Nevertheless, the above potential is defined on general parameters whose adjustment equals picking different supergravity backgrounds. Therefore, the failure of Kovacic's algorithm here just states that \textit{not all} possible backgrounds are integrable. It does not say that there are no integrable ones, among the whole class. This can be also realized by the fact that we have already found, in the previous subsection \ref{NATDclassI}, an integrable selection of parameters, i.e. $c_1=c_4=0$. It's this kind of possible combinations of these parameters (like $c_1=c_4=0$), i.e. particular supergravity backgrounds, that we are interested in, if any (others) exist.\\

Therefore, we shall utilize the full power of Kovacic's method. This way, if there are any selections of $\lbrace c_i,F_0\rbrace$ that allow for Liouvillian solutions of (\ref{explicitNVEr}), we shall find them along with their associated solutions. If such selections are impossible, then we shall \textit{safely} declare the whole supergravity class as non-integrable.\\

Kovacic's analytic algorithm is a step-by-step procedure, detailed in Appendix \ref{appendixA}. Overall, it states that each one of its criteria is associated with a sub-algorithm, called a Case, that may (or may not) solve the equation at hand. As proved above, our NVE (\ref{explicitNVEr}) satisfies the first and second criteria and, thus, must be undertaken by Cases 1 and 2, respectively, of the algorithm.\\

Since there is nothing intuitive about Kovacic's method, the explicit calculations of the analytic algorithm on all Cases are held in Appendix \ref{appendixB}. On the main article, we just present the results of the algorithm and act with our string theory considerations on them.\\

\subsubsection{Case 1}\label{classIcase1}
Case 2 takes into account that Case 1 does not hold, hence we shall always begin by considering Case 1 of Kovacic's theorem. The algorithm for this particular Case is explained in Appendix \ref{case1algorithm} and the explicit calculation on our $r$ NVE (\ref{explicitNVEr}) is given in Appendix \ref{case1classI}.\vspace{0.2cm}

Up to some real constants and signs that we do not care about here, the algorithm produces the quantity

\begin{equation}
d\:\sim\: i\sqrt{\frac{c_1c_4}{c_3^2}}\pm\frac{i\,(c_1c_5+c_4F_0)E}{2\sqrt{c_3^2c_5F_0E^2+c_3^4\,\nu^2}}\label{dforCase1r}
\end{equation}\\
and states that $d$ has to be a \textit{non-negative integer}. If $d$ is such a number, then the algorithm moves on to its next stage. If $d$ is never such a number, then Case 1 cannot give a Liouvillian solution. In other words, integrability demands the above object to be real.\vspace{0.3cm}

Therefore, we have reduced our integrability problem to whether there are any inter-relations between the supergravity parameters $\lbrace c_i,F_0\rbrace$ that let (\ref{dforCase1r}) to be real. Such a relation would correspond to a specific background. In what follows, we prove that these parameters are constrained by the behavior of the rank functions $h_4(\rho\,;c_4,c_5)$ and $h_8(\rho\,;c_1,F_0)$, in such a way that no such relations can exist.\vspace{0.3cm}

So, there are three possibilities for (\ref{dforCase1r}) to be real: either both imaginary terms vanish simultaneously, either they cancel each other out or they both end up real.\vspace{0.3cm}

The first possibility is excluded since $c_1,c_4\neq0$, the opposite being true only on the first $\rho$ interval of the space (the NATD part). Alternatively, if $c_1=0$ while $c_4\neq0$ then the first term may vanish but the second one (which also has to vanish) implies $F_0=0$, which together lead to $h_8=0$. But, as argued repeatedly, $h_4,h_8=0$ can only happen at the beginning and at the end of the space, otherwise the background degenerates and blows up. The same holds for $c_1\neq0$ while $c_4=0$.\vspace{0.3cm}

The second possibility is also excluded, since the first term in $\nu$-independent and the second $\nu$-dependent. $\nu$ is the string winding number and can be anything, while we want a relation between parameters for all possible string configurations. Notice that this is another good example of why all the available stringy character, in a supergravity (non-) integrability test, is always welcome.\vspace{0.3cm}

Last but not least, the third possibility is excluded too, since in (\ref{c1c4>0}) we proved that $c_1c_4\geq0$ always and, hence, the first term in (\ref{dforCase1r}) can never be a positive real number. Since the first term cannot be real nor vanish we don't have to check whether the second term does.\vspace{0.3cm}

Nevertheless, let us look up the second term of (\ref{dforCase1r}), for completeness. The second term has a $\nu$-dependent square root, meaning that the root argument cannot be fixed as negative and, thus, cannot produce an $i$ factor in order to end up with a non-zero real number. Therefore, the only possibility left is for this term to vanish. This only happens when

\begin{equation}
c_1c_5\:=\:-c_4F_0\hspace{1cm}\Rightarrow\hspace{1cm} c_1\:=\:-\frac{c_4F_0}{c_5}
\end{equation}\\
which, if we substitute in the first term of (\ref{dforCase1r}) and demand reality, gives $c_5F_0>0$. But then, given that $c_5F_0>0$ together with $c_1c_4>0$, the initial assumption $c_1c_5=-c_4F_0$ can never hold\footnote{We can include the possibility that $c_1c_5+c_4F_0=0$ when $c_5=F_0=0$, but then this doesn't stop the first term from being imaginary.}. As expected, we end up with the same result.\vspace{0.3cm}

One could also argue whether the instantonic mode $E=0$ is an option to vanish the second term in (\ref{dforCase1r}). The fact is that by choosing $E=0$, we select a particular configuration for our embedding. Even if the $E=0$ mode was integrable it would make no difference, since for $E\neq0$ the configurations are non-integrable as shown above. While an integrable sector of the theory should exhibit its homonymous property on its wholeness, i.e. for all configurations of the string embedding. That is the reason we only look for special selections of $\lbrace c_i,F_0\rbrace$, but not of $E,\nu,\kappa$. For the curious mind, the instanton $E=0$ leads here to a non-Liouvillian solution.\\

Subsequently, $d$ can never be a non-negative integer and, thus, Case 1 cannot provide us a Liouvillian solution. Of course, our NVE (\ref{explicitNVEr}) also satisfies the second Kovacic's criterion and, to that end, we still have a chance to spot integrability through Case 2.\\

\subsubsection{Case 2}\label{classIcase2}
This Case is explained in Appendix \ref{case2algorithm} and the explicit calculation on our $r$ NVE (\ref{explicitNVEr}) is given in Appendix \ref{case2classI}. Here, the algorithm produces the integer quantities $E_i\cap\mathbb{Z}$,

\begin{equation}
E_1\:=\:\left\lbrace2-4\sqrt{-\frac{c_1c_4}{c_3^2}}\,,\;2\;,\,2+4\sqrt{-\frac{c_1c_4}{c_3^2}}\right\rbrace\hspace{1.5cm}E_2\:=\:E_3\:=\:\lbrace-1,2,5\rbrace
\end{equation}\\
However, as already shown in (\ref{c1c4>0}) and used on the previous Case, $c_1c_4\geq0$. Which means that the quantities under the square roots in $E_1$ are non-positive and thus give overall imaginary numbers or 2. In any case, since $E_i$'s have to be integers, we conclude that $E_1=\lbrace2\rbrace$.\\

Given these $E_i$'s, the algorithm builds a rational function based on the pole structure of $\mathcal{V}_\varrho$ as

\begin{equation}
\theta\:=\:\frac{1}{\tau}-\frac{1}{2\left(\tau+\frac{c_4}{c_5E}\right)}-\frac{1}{2\left(\tau+\frac{c_1}{F_0E}\right)}\label{thetaCase2ClassI}
\end{equation}\\
and dictates that the equation

\begin{equation}
\theta''+3\theta\theta'+\theta^3-4\mathcal{V}_\varrho\theta-2\mathcal{V}_\varrho'=0
\end{equation}\\
must be satisfied, in order for a Liouvillian solution to exist. Replacing $\theta$, (\ref{thetaCase2ClassI}), into the latter necessary condition we find out that it is not satisfied. Therefore, Case 2 also fails to provide a Liouvillian solution.\\

Since both Cases failed to expose integrability, we may now declare this class of supergravity backgrounds as non-integrable. Of course, since dynamics along the $r$ dimension is non-integrable we don't have to study the NVE for $\chi$ and our analysis can cease at this point.\\

This whole section, dedicated on the $r$ NVE (\ref{explicitNVEr}), was a prototype example of the analytic enforcement of Kovacic's algorithm. Since this differential equation was parametrized by $\lbrace c_1,..,c_5,F_0\rbrace$ we employed the algorithm analytically in order to find any special relations between the parameters that would allow for a Liouvillian solution. In our particular case, however, by demanding consistency on those brane set-up parameters, we proved that no such relations can exist.\newpage

The bottom line is that the above procedure is necessary if one wants to study non-integrability, through differential Galois theory, on a parametrized differential equation. Failure of Kovacic's algorithm without exploring the possible selections between the parameters does not imply the non-integrability of the system. It just states that not all choices of the parameters lead to an integrable system. By which we mean that particular combinations of the parameters may produce Liouvillian solutions. That is, if we are allowed to play with the parameters. If full generality on them is necessary, for any reason, then the analytic application of the algorithm is not needed.\\

\section{Class II backgrounds}\label{ClassIIsection}
Reminding ourselves for one last time the general form of the NS-NS sector of the AdS$_3\,\times\,$S$^2\times\mathbb{R}\,\times\,$CY$_2$ massive IIA supergravity, at string frame, as
\begin{equation}
\begin{split}
\dd s^2\:=\:f_1\,\dd s^2_{\mbox{\tiny AdS$_3$}}+f_2\,\dd s^2_{\mbox{\tiny S$^2$}}+\frac{\dd\rho^2}{f_1}+f_3\,\dd s^2_{\mbox{\tiny CY$_2$}}\hspace{2cm}\\[10pt]
B_2\:=\:f_4\,\mbox{vol}_{\mbox{\scriptsize S$^2$}}\hspace{2cm}e^{-\Phi}\:=\:f_5\hspace{2cm}f_i\:=\:f_i\left(u,h_4,h_8\right)\label{GeneralAdS3background3}
\end{split}
\end{equation}\\
then the second supergravity class is defined by the warp factors

\begin{equation}
\begin{split}
f_1\:=\:\frac{u}{\sqrt{h_4^2-h_8^2}}\hspace{1.5cm}f_2\:=\:f_1\frac{h_4^2-h_8^2}{4(h_4^2-h_8^2)+(u')^2}\hspace{1.5cm}f_3\:=\:\frac{\sqrt{h_4^2-h_8^2}}{h_4}\hspace{0.8cm}\\[10pt]
f_4\:=\:\frac{1}{2}\left(-\rho+\frac{uu'}{4(h_4^2-h_8^2)+(u')^2}\right)+\frac{h_8}{h_4}\hat{J}\hspace{2cm}f_5\:=\:\frac{h_4\sqrt{4(h_4^2-h_8^2)+(u')^2}}{2\sqrt{u}(h_4^2-h_8^2)^{\frac{1}{4}}}
\end{split}\label{ClassIIwarps}
\end{equation}\\
where $\hat{J}$ is a 2-form on CY$_2$. For simplicity, again, we treat the functions $h$, $u$ in a general manner, as in (\ref{uh4h8general}), i.e.

\begin{equation}
u(\rho)\:=\:c_3\rho\hspace{1.3cm}h_4(\rho)\:=\:c_4+c_5\rho\hspace{1.3cm}h_8(\rho)\:=\:c_1+F_0\rho
\end{equation}\\
since their piecewise character, (\ref{h4piecewise})-(\ref{h8piecewise}), can be always assumed. Observe that it must be always true that $h_4\geq h_8\geq0$.\\

\begin{figure}[h!]
    \centering
    {{\includegraphics[width=9cm]{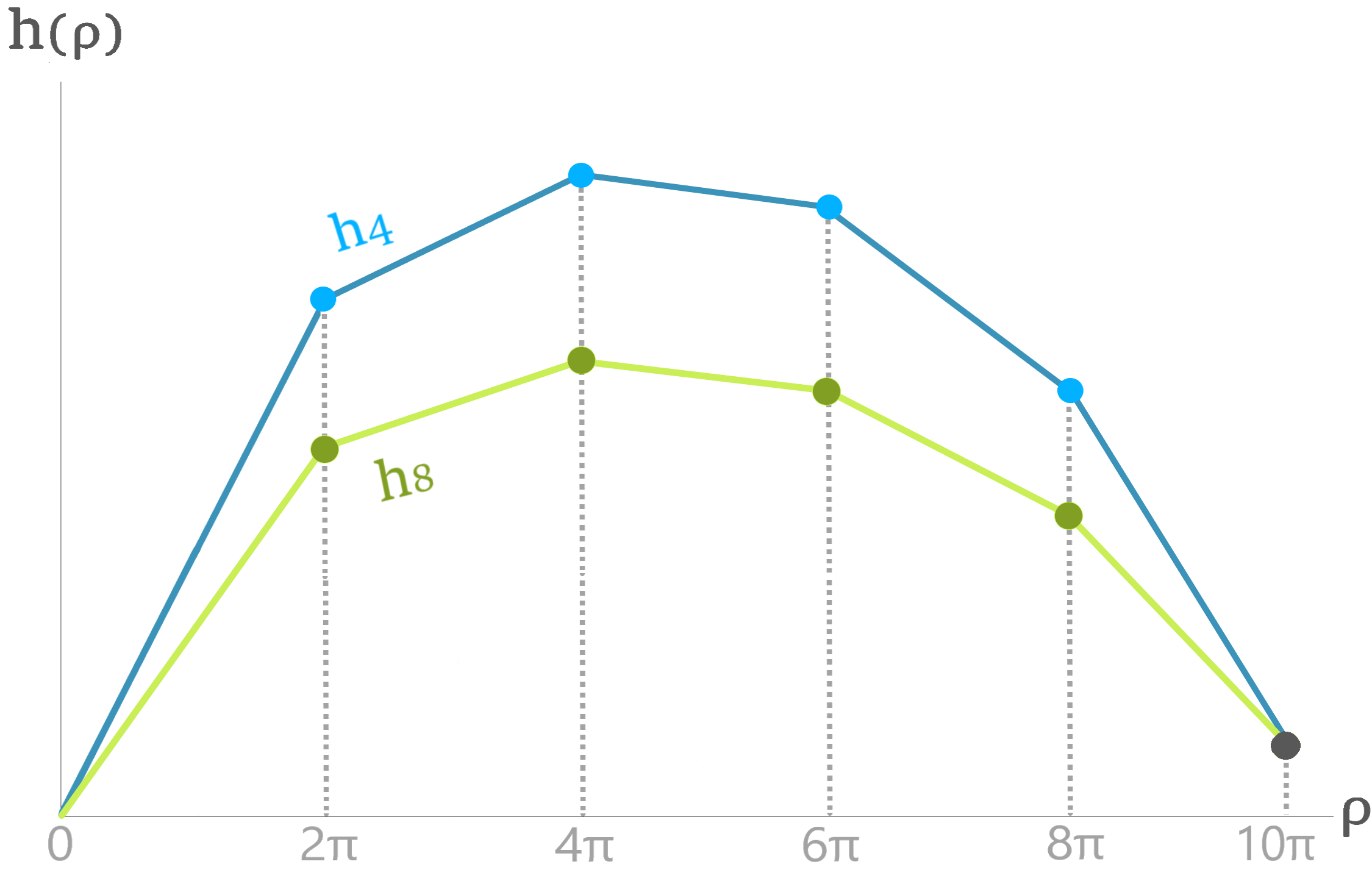} }}%
\caption{An example of the linear functions $h_{4,8}(\rho)$ in Class II supergravity. These piecewise functions start from $h_4|_{\rho=0}=h_8|_{\rho=0}=0$, with $h_4>h_8$ always, and decrease in slope until they reunite at the end, $\rho_f=10\pi$.}
\label{figure3}
\end{figure}

Notice that, in this supergravity class, the condition for the background to be smooth at the beginning and at the end of the $\rho$ dimension is $h_4|_{\rho=0}=h_8|_{\rho=0}=0$ and $h_4|_{\rho_f}=h_8|_{\rho_f}$, respectively. Hence, $h_4$ and $h_8$ are positive piecewise linear curves that start from $h_4|_{\rho=0}=h_8|_{\rho=0}=0$, with $h_4>h_8$ always, and decrease in slope until they reunite at the end, $\rho_f$, as in Figure \ref{figure3}.
\\

\subsection{NVE for $r$}
Faithful to the way we treated Class I, let us begin our integrability analysis by first studying the string dynamics along $r$. We again replace the warp factors (\ref{ClassIIwarps}) into the NVE for $r$, (\ref{newNVEr}), and obtain

\begin{equation}
\ddot{\varrho}\:=\:\frac{Q_{\mbox{\tiny II}}}{\tau^2(\tau-\frac{c_1-c_4}{(c_5-F_0)E})^2(\tau+\frac{c_1+c_4}{(c_5+F_0)E})^2}\,\varrho\label{explicitNVErII}
\end{equation}\\
\\
where $Q_{\mbox{\tiny II}}=Q_{\mbox{\tiny II}}(\tau^6,c_i,F_0,E)$ is a long polynomial in the numerator whose explicit form will not concern us. In this class, $\mathcal{V}_\varrho$ also comes with three poles of order two, $\lbrace\tau_1=0,\tau_2=\frac{c_1-c_4}{(c_5-F_0)E},\tau_3=-\frac{c_1+c_4}{(c_5+F_0)E}\rbrace$ and it expands around $\tau\rightarrow\infty$ as
\begin{equation}
\mathcal{V}_\varrho^\infty\:=\:-\left(\frac{(c_5^2-F_0^2)E^2}{c_3^2}+\nu^2\right)-\frac{2(c_4c_5-c_1F_0)E}{c_3^2\tau}+\mathcal{O}\left(\frac{1}{\tau^2}\right)\label{VrinfinityII}
\end{equation}\\
exhibiting zero order behavior there. Thus, $\mathcal{V}_\varrho$ satisfies the first and second Kovacic's criteria, implying that the NVE (\ref{explicitNVErII}) may have Liovillian solutions. However, Kovacic's algorithm fails in this class too to solve it as it is.\vspace{0.3cm}

Of course, the NVE (\ref{explicitNVErII}) is again parametrized by $\lbrace c_i,F_0\rbrace$, whose various inter-relations give different backgrounds in this supergravity class. Therefore, we shall employ for one last time the full power of Kovacic's method to seek out for any such relations that allow for Liouvillian solutions, if any.\vspace{0.3cm}

Since in this class, the $r$ NVE (\ref{explicitNVErII}) satisfies the first and second Kovacic's criteria too, we will again consider Cases 1 and 2 of Kovacic's theorem.

\subsubsection{Case 1}\label{classIIcase1}
As said before, Case 2 takes into account that Case 1 does not hold, thus we again begin by considering Case 1 of Kovacic's theorem. The explicit calculation on our $r$ NVE (\ref{explicitNVErII}) is given in Appendix \ref{case1classII}. Here, up to some real constants and signs, the algorithm produces the quantity

\begin{equation}
d\:\sim\: i\sqrt{\frac{c_4^2-c_1^2}{c_3^2}}\pm\frac{i\,(c_4c_5-c_1F_0)E}{\sqrt{c_3^2(c_5^2-F_0^2)E^2+c_3^4\,\nu^2}}\label{dforCase1rII}
\end{equation}\\
Again, $d$ has to be a non-negative integer for Case 1 to produce a Liouvillian solution, which in turn means that the above object must be real.\vspace{0.3cm}

The history repeats itself. There are three possibilities for (\ref{dforCase1rII}) to be real: either both imaginary terms vanish simultaneously, either they cancel each other out or they both end up real. Considering the $\nu$-dependence of the second term, that term can never be a non-zero real number since $\nu$ can be anything for a general string configuration. On the exact same grounds, it can never be canceled against the first term, which is $\nu$-independent. Those arguments exclude the second and third possibility.\vspace{0.3cm}

The only possibility left is for the second term of (\ref{dforCase1rII}) to vanish, i.e. $c_4c_5=c_1F_0$. In turn, the latter condition obligates the first term to give $\abs{c_5}\geq\abs{F_0}$, in the name of reality.\vspace{0.3cm}

Now, as we argued in the beginning of the section and showed in Figure \ref{figure3}, $h_4$ and $h_8$ are positive piecewise curves that both start from $h_4|_{\rho=0}=h_8|_{\rho=0}$ with $h_4>h_8$ everywhere, and decrease in slope until they reunite at the end, $h_4|_{\rho_{f}}=h_8|_{\rho_{f}}$. From simple trigonometry, the fact that $h_4$ is always above $h_8$ while they both end at the same point $\rho_f$ states that: \textit{at least} on the last interval before their reunion, it is true that $c_4>c_1$. Whatever their slope inter-relation is. Observing Figure \ref{figure3}, this statement is equivalent to saying that, on the last interval, $h_4$ always has a greater projection on the $\rho=0$ axis than $h_8$.\vspace{0.3cm}

But now, since there has to be at least one region where $c_4>c_1$, then, combined with the hypothesis $\abs{c_5}\geq\abs{F_0}$, the initial assumption $c_4c_5=c_1F_0$ can never hold everywhere.\vspace{0.3cm}

Therefore, $d$ can never be a non-negative integer and we conclude that Case 1 fails to provide a Liouvillian solution for the second supergravity class. Since $\mathcal{V}_\varrho$ satisfies also the second Kovacic's criterion, we move on to examine whether Case 2 can do any better.

\subsubsection{Case 2}\label{classIIcase2}
For this last application of Case 2 in Kovacic's theorem, the explicit calculation on our $r$ NVE (\ref{explicitNVErII}) is given in Appendix \ref{case2classII}. Here, the algorithm produces the integer quantities $E_i\cap\mathbb{Z}$,
\begin{equation}
E_1\:=\:\left\lbrace2-4\sqrt{\frac{c_1^2-c_4^2}{c_3^2}}\,,\;2\;,\,2+4\sqrt{\frac{c_1^2-c_4^2}{c_3^2}}\right\rbrace\hspace{1.5cm}E_2\:=\:E_3\:=\:\lbrace-1,2,5\rbrace
\end{equation}\\
However, as we just showed on the previous subsection, $c_4>c_1$ at least at the last interval before $h_4$ and $h_8$ meet at $\rho_f$. Thus $c_1\geq c_4$ can never be always true for any interval, which means that the square root in $E_1$ becomes imaginary. Hence, since $E_i$'s have to be integers, we conclude that $E_1=\lbrace2\rbrace$.\\

Since the $E_i$'s are exactly the same with the ones of Class I, the algorithm again builds a rational function, based on the poles of $\mathcal{V}_\varrho$ in Class II, as

\begin{equation}
\theta\:=\:\frac{1}{\tau}-\frac{1}{2\left(\tau-\frac{c_1-c_4}{(c_5-F_0)E}\right)}-\frac{1}{2\left(\tau+\frac{c_1+c_4}{(c_5+F_0)E}\right)}\label{thetaCase2ClassII}
\end{equation}\\
and, the same as the last time, dictates that the equation $\theta''+3\theta\theta'+\theta^3-4\mathcal{V}_\varrho\theta-2\mathcal{V}_\varrho'=0$ should be satisfied. In this class too it does not, therefore Case 2 cannot provide us a Liouvillian solution either, for our $r$ NVE (\ref{explicitNVErII}).\\

Since both Cases also failed for this class of backgrounds, for any possible selection of the parameters $\lbrace c_i,F_0\rbrace$, we declare this supergravity family too as non-integrable. Hence, both supergravity classes are non-integrable and that concludes our integrability adventure on this AdS$_3$ supergravity.\\

\section{Epilogue}\label{epilogue}
The apparent conclusion of the present work is the complete, classical, Liouvillian non-integrability on certain warped backgrounds of the form AdS$_3\,\times\,$S$^2\times\mathbb{R}\,\times\,$CY$_2$, first constructed in \cite{Lozano:2019emq} and then considered in \cite{Lozano:2019zvg}. Enforcing the full power of Kovacic's theorem, along with simple consistency considerations on the supergravity brane set-ups, we deduced that all possible backgrounds in this warped AdS$_3$ supergravity family are non-integrable.\\

Note that those considerations were not based on the supergravity approximation of the parameters of the background, which would be an easier but less general way to go. Instead we considered the consistency rules of string theory on Hanany-Witten brane set-ups.\\

An exception of two integrable choices of backgrounds is when the Class I supergravity solution reduces to the ATD and NATD of AdS$_3\,\times\,$S$^3\times\,$T$^4$, for all intervals along the $\rho$ dimension. These unique integrable cases occur when AdS$_3$ unwarps from the rest of the space. Any other warped background for both AdS$_3$ supergravity classes, was proven to be non-integrable.\\

As a side comment, we note that integrability on AdS supergravity vacua seems to occur only when the AdS part of the space gets unwarped. In the present case, we illustrated that this only happens on Class I, when the background reduces to the integrable ATD and NATD of AdS$_3\,\times\,$S$^3\times\,$T$^4$. Then, there is the Sfetsos-Thompson background \cite{Sfetsos:2010uq,Nunez:2018qcj}, which is the unwarped integrable case of the Gaiotto-Maldacena AdS$_5$ vacua, \cite{Gaiotto:2009gz}. The same also holds for a more recent background \cite{Filippas:2019puw}, among the AdS$_7$ massive IIA supergravity family \cite{Apruzzi:2013yva,Apruzzi:2015wna}. This argument still holds as just a dominant indication and certainly not as definite statement. However, in \cite{Wulff:2019tzh} and later in \cite{Giataganas:2019xdj}, it was illustrated that on AdS supergravity vacua that allow for the GKP embedding the AdS space should be unwarped for integrability to occur. This constitutes a strong constraint for many AdS backgrounds, yet it does not apply in our AdS$_3$ family which does not support a GKP vacuum.\\

Nevertheless, the main aspect of this work is the way we utilize Kovacic's theorem on a differential equation. We illustrated that failure of Kovacic's algorithm on a parametrized equation does not necessarily imply absence of Liouvillian solutions. It just says that there are no such solutions for the \textit{full generality} of the parameters. If the problem allows to impose any restrictions on its parameters, then a brand new horizon of possibilities appears. On the other hand, if full generality on them is necessary, for any reason, then the analytic application of the algorithm is not needed. In the case when the parameters are adjustable, like with our present supergravity family, then the analytic algorithm \textit{must} be employed. This way, if there are any selections between the parameters that lead to an integrable result, the algorithm will find them along with the corresponding solutions. Only when this procedure is followed and no such selections are discovered, then we can \textit{safely} deduce that our system is non-integrable in the Liouvillian sense.\\

In our case, the AdS$_3$ supergravity family is defined on general parameters whose adjustment equals picking different supergravity backgrounds. Therefore, the failure of Kovacic's algorithm here just states that \textit{not all} possible backgrounds are integrable. It does not say that there are no integrable ones, among the whole family. Therefore, we utilized the full power of Kovacic's theorem, by considering its analytic algorithm, and found some necessary conditions $-$ on the background parameters $-$ in order for Liouvillian solutions to exist. By constraining these parameters according to the consistency of the associate brane set-ups, we proved that those necessary conditions can never hold, yielding the complete non-integrability of these vacua. That is, up to the trivial case where the background reduces to the ATD and NATD of AdS$_3\,\times\,$S$^3\times\,$T$^4$.\\
\\


\paragraph{Acknowledgments}
I thank C. Nunez for various discussions on this work. I also thank K. Sfetsos, T. Hollowood, N. Macpherson and D. Thompson for helpful comments and D. Giataganas for an enlightening discussion on the draft of the paper. This work is supported by an STFC scholarship.\\

\appendix

\section{Differential Galois theory and Kovacic's theorem}\label{appendixA}
In this appendix we give the basic elements of differential Galois theory that were used by Kovacic \cite{Kovacic} to produce his infamous algorithm, regarding the existence of Liouvillian solutions on second order linear ordinary differential equations. By a Liouvillian, closed form solution we mean one that is given in terms of algebraic, exponential, trigonometric functions and integrals of those. \\

The theorem concerns second order linear ordinary differential equations of the form

\begin{equation}
y''(x)+\mathcal{B}(x)y'(x)+\mathcal{A}(x)y(x)\:=\:0
\end{equation}\\
where $x\in\mathbb{C}$ and $\mathcal{A}\mathcal{B}$ are rational complex functions. We can use the variable transformation $y=e^{\frac{1}{2}\int\mathcal{B}}z$ to eliminate the $y'$ term and acquire the new equation

\begin{equation}
z''(x)\:=\:\mathcal{V}(x)\,z(x)\hspace{2cm}\mathcal{V}=\frac{1}{4}\left(2\mathcal{B}'+\mathcal{B}^2-4\mathcal{A}\right)\label{transformedNVEappendix}
\end{equation}\\
where we shall call $\mathcal{V}$ the potential of the differential equation. Evidently, $y$ exhibits Liouvillian solutions if and only if $z$ does, thus no generality is lost through this change of variable.\\

The starting point of differential Galois theory on this kind of equations, which is actually Piccard-Vessiot theory, is the group of automorphisms of its solutions, that is SL$(2,\mathbb{C})$ and its possible subgroups. Letting $G$ be an algebraic subgroup of SL$(2,\mathbb{C})$, then one of the four cases can occur:\\

\case{1} \textit{$G$ is triangulisable.}\\

\case{2} \textit{$G$ is conjugate to a subgroup of}

\begin{equation}
\left\lbrace\left.\left(\begin{array}{cc}
c &0\\
0 &c^{-1}
\end{array}\right)\right|c\in\mathbb{C}, c\neq0\right\rbrace\cup\left\lbrace\left.\left(\begin{array}{cc}
0 &c\\
-c^{-1} &0
\end{array}\right)\right|c\in\mathbb{C}, c\neq0\right\rbrace
\end{equation}
\hspace{2.1cm} \textit{and Case 1 does not hold.}\\

\case{3} \textit{$G$ is finite and Cases 1 and 2 do not hold.}\\

\case{4} \textit{$G=$ SL$(2,\mathbb{C})$.}\\

If the differential equation falls into one of the three first cases, it has Liouvillian solutions. On the other hand, if $G=$ SL$(2,\mathbb{C})$, no such solutions can exist.\\

The first contribution by Kovacic was to translate Cases 1, 2 and 3 into algebraic arguments on the behavior of $\mathcal{V}$ in (\ref{transformedNVEappendix}). These algebraic conditions build up the following theorem.\\

\theorem{} \textit{The following conditions are necessary for the respective Cases to hold.}\\

\case{1} \textit{Every pole of $\mathcal{V}$ must have even order or else have order 1. The order of $\mathcal{V}
$ at $\infty$ must be even or else greater than 2.}\\

\case{2} \textit{$\mathcal{V}$ must have at least one pole that either has odd order greater than 2 or else has order 2.}\\

\case{3} \textit{The order of a pole of $\mathcal{V}$ cannot exceed 2 and the order of $\mathcal{V}$ at $\infty$ must be at least 2.}\\

If $\mathcal{V}=s/t$, then the poles of $\mathcal{V}$ are the zeros of $t$ and the order of the pole is the multiplicity of the zero of $t$. By the order of $\mathcal{V}$ at $\infty$ we shall mean the number $\deg t-\deg s$.\\

Since these conditions are \textit{necessary} for the respective cases to hold, then also their failure is \textit{sufficient} for Case 4 to hold. Therefore we deduce that failure of all three conditions is enough to declare the differential equation (\ref{transformedNVEappendix}) as non-integrable in the Liouvillian sense.\\

Nevertheless, if any of the conditions is satisfied, then the respective Case may hold and if it does then a Liovillian solution exists. Hence, when a condition is satisfied we are prompted to the sub-algorithm of the respective Case to examine whether such a solution exists and, when it does, use the algorithm to find it. The second contribution by Kovacic was to produce these algorithms for Cases 1, 2 and 3.\\

\subsection{The algorithm for Case 1}\label{case1algorithm}
We assume that the necessary condition of Case 1 holds, and we denote by $\Gamma$ the set of poles of $\mathcal{V}$.\\

\step{1} For each $c\in\Gamma\cup\lbrace\infty\rbrace$ we define a rational function $[\sqrt{\mathcal{V}}]_c$ and two complex
numbers $\alpha_c^\pm$ as described below.\\

\cc{1} If $c\in\Gamma$ and $c$ is a pole of order 1, then\\
\begin{equation*}
[\sqrt{\mathcal{V}}]_c\:=\:0\hspace{2cm}\alpha_c^\pm\:=\:1
\end{equation*}

\cc{2} If $c\in\Gamma$ and $c$ is a pole of order 2, then\\
\begin{equation*}
[\sqrt{\mathcal{V}}]_c\:=\:0
\end{equation*}\\
Let $\beta_c$ be the coefficient of $1/(x-c)^2$ in the partial fraction expansion for $\mathcal{V}$. Then\\
\begin{equation*}
\alpha_c^\pm\:=\:\frac{1}{2}\pm\frac{1}{2}\sqrt{1+4\beta_c}
\end{equation*}

\cc{3} If $c\in\Gamma$ and $c$ is a pole of order $2\nu\geq4$ (necessarily even by the condition for Case 1), then $[\sqrt{\mathcal{V}}]_c$ is the sum of terms involving $1/(x-c)^i$ for $2\leq i\leq\nu$ in the Laurent series expansion of $\sqrt{\mathcal{V}}$ at $c$. There are two possibilities for $[\sqrt{\mathcal{V}}]_c$, one being the negative of the other, either one may be chosen. Thus\\
\begin{equation*}
[\sqrt{\mathcal{V}}]_c\:=\:\frac{a}{(x-c)^\nu}+\dots+\frac{d}{(x-c)^2}
\end{equation*}\\
Let $\beta_c$ be the coefficient of $1/(x-c)^{\nu+1}$ in $\mathcal{V}$ minus the coefficient of $1/(x-c)^{\nu+1}$ in $[\sqrt{\mathcal{V}}]^2_c$. Then\\
\begin{equation*}
\alpha_c^\pm\:=\:\frac{1}{2}\left(\pm\frac{\beta_c}{a}+\nu\right)
\end{equation*}\\

\cci{1} If the order of $\mathcal{V}$ at $\infty$ is $>2$, then\\
\begin{equation*}
[\sqrt{\mathcal{V}}]_\infty\:=\:0\hspace{2cm}\alpha_\infty^+\:=\:0\hspace{1cm}\alpha_\infty^-\:=\:1
\end{equation*}

\cci{2} If the order of $\mathcal{V}$ at $\infty$ is 2, then\\
\begin{equation*}
[\sqrt{\mathcal{V}}]_\infty\:=\:0
\end{equation*}\\
Let $b_\infty$ be the coefficient of $1/x^2$ in the Laurent series expansion of $\mathcal{V}$ at $\infty$. (If $\mathcal{V}=s/t$, where $s$, $t$ are relatively prime, then $b_\infty$ is the leading coefficient of $s$ divided by the leading coefficient of $t$.) Then\\
\begin{equation*}
\alpha_\infty^\pm\:=\:\frac{1}{2}\pm\frac{1}{2}\sqrt{1+4\beta_\infty}
\end{equation*}

\cci{3} If the order of $\mathcal{V}$ at $\infty$ is $-2\nu\leq0$ (necessarily even by the condition of Case 1), then $[\sqrt{\mathcal{V}}]_\infty$ is the sum of terms involving $x^i$ for $0\leq i\leq\nu$ in the Laurent series for $\sqrt{\mathcal{V}}$ at $\infty$. (Either one of the two possibilities may be chosen.) Thus\\
\begin{equation*}
[\sqrt{\mathcal{V}}]_\infty\:=\:ax^\nu+\dots+d
\end{equation*}\\
Let $\beta_\infty$ be the coefficient of $x^{\nu-1}$ in $\mathcal{V}$ minus the coefficient of $x^{\nu-1}$ in $([\sqrt{\mathcal{V}}]_\infty)^2$. Then\\
\begin{equation*}
\alpha_\infty^\pm\:=\:\frac{1}{2}\left(\pm\frac{\beta_\infty}{a}-\nu\right)
\end{equation*}\\

\step{2} For each family $s=(s(c))_{c\in\Gamma\cup\lbrace\infty\rbrace}$, where $s(c)$ is $+$ or $-$, let\\
\begin{equation*}
d\:=\:\alpha_\infty^{s(\infty)}-\sum_{c\in\Gamma}\alpha_c^{s(c)}
\end{equation*}\\
If $d$ is a non-negative integer, then\\
\begin{equation*}
\omega\:=\:\sum_{c\in\Gamma}\left(s(c)[\sqrt{\mathcal{V}}]_c+\frac{\alpha_c^{s(c)}}{x-c}\right)+s(\infty)[\sqrt{\mathcal{V}}]_\infty
\end{equation*}\\
is a candidate for $\omega$. If $d$ is not a non-negative integer, then the family $s$ may be removed from consideration.\\

\step{3} This step should be applied to each of the families retained from Step 2, until success is achieved or the supply of families has been exhausted. In the latter event, Case 1 cannot hold.\\[5pt]
For each family, search for a monic polynomial $P$ of degree $d$ (as defined in Step 2) that satisfies the differential equation\\
\begin{equation*}
P''+2\omega P'+(\omega'+\omega^2-\mathcal{V})P\:=\:0
\end{equation*}\\
This is conveniently done by using undetermined coefficients and is a simple problem in linear algebra, which may or may not have a solution. If such a polynomial exists, then $\eta=Pe^{\int\omega}$ is a solution of the differential equation (\ref{transformedNVEappendix}). If no such polynomial is found for any family retained from Step 2, then Case 1 cannot hold.\\

\subsection{The algorithm for Case 2}\label{case2algorithm}
This algorithm assumes that Case 1 is known to fail. Just as for Case 1, we first collect data for each pole $c$ of $\mathcal{V}$ and also for $\infty$. The form of the data is a set $E_c$ (or $E_\infty$) consisting of from one to three integers. Next we consider families of elements of these sets, perhaps discarding some and retaining others. If no families are retained, Case 2 cannot hold. For each family retained we search for a monic polynomial that satisfies a certain linear differential equation. If no such polynomial exists for any family, then Case 2 cannot hold. If such a polynomial does exist, then a solution to the differential equation (\ref{transformedNVEappendix}) has been found.\\

Let $\Gamma$ be the set of poles of $\mathcal{V}$.\\

\step{1} For each $c\in\Gamma$ we define $E_c$ as follows.\\

\cc{1} If $c$ is a pole of order 1, then $E_c=\lbrace4\rbrace$.\\

\cc{2} If $c$ is a pole of order 2 and if $\beta_c$ is the coefficient of $1/(x-c)^2$ in the partial fraction expansion of $\mathcal{V}$, then
\begin{equation*}
E_c=\lbrace 2+k\sqrt{1+4\beta_c}|k=0,\pm2\rbrace\cap\mathbb{Z}
\end{equation*}

\cc{3} If $c$ is a pole of order $\nu>2$, then $E_c=\lbrace\nu\rbrace$.\\

\cci{1} If $\mathcal{V}$ has order $>2$ at $\infty$ , then $E_\infty=\lbrace0,2,4\rbrace$.\\

\cci{2} If $\mathcal{V}$ has order 2 at $\infty$ and $\beta_\infty$ is the coefficient of $\mathcal{V}$ in the Laurent series expansion of $\mathcal{V}$ at $\infty$, then
\begin{equation*}
E_\infty=\lbrace 2+k\sqrt{1+4\beta_\infty}|k=0,\pm2\rbrace\cap\mathbb{Z}
\end{equation*}

\cci{3} If the order of $\mathcal{V}$ at $\infty$ is $\nu<2$, then $E_\infty=\lbrace\nu\rbrace$.\\

\step{2} We consider all families $(e_c)_{c\in\Gamma\cup\lbrace\infty\rbrace}$ with $e_c\in E_c$. Those families all of whose coordinates are even may be discarded. Let\\
\begin{equation*}
d\:=\:\frac{1}{2}\left(e_\infty-\sum_{c\in\Gamma}e_c\right)
\end{equation*}\\
If $d$ is a non-negative integer, the family should be retained, otherwise the family is discarded. If no families remain under consideration, Case 2 cannot hold.\\

\step{3} For each family retained from Step 2, we form the rational function\\
\begin{equation*}
\theta\:=\:\frac{1}{2}\sum_{c\in\Gamma}\frac{e_c}{x-c}
\end{equation*}\\
Next we search for a monic polynomial $P$ of degree $d$ (as defined in Step 2) such that\\
\begin{equation*}
P'''+3\theta P''+(3\theta^2+3\theta'-4\mathcal{V})P'+(\theta''+3\theta\theta'+\theta^3-4\mathcal{V}\theta-2\mathcal{V}')P\:=\:0
\end{equation*}\\
If no such polynomial is found for any family retained from Step 2, then case 2 cannot hold.\\[5pt]
Suppose that such a polynomial is found. Let $\varphi=\theta+P'/P$ and let $\omega$ be a solution of the equation
\begin{equation*}
\omega^2+\varphi\omega+(\frac{1}{2}\varphi'+\frac{1}{2}\varphi^2-\mathcal{V})\:=\:0
\end{equation*}\\
Then $\eta=e^{\int\omega}$ is a solution of the differential equation (\ref{transformedNVEappendix}).\\

We will not go on to describe the algorithm for Case 3, since we will not be needing it on the present analysis, while it is a bit more of a job than the above Cases 1 and 2. We should note, however, that the necessary algebraic condition that allows for Case 3 to hold is quite restricting and certainly more rare than the others to its satisfaction. If the reader still desires the explicit sub-algorithm for Case 3, Kovacic's original work \cite{Kovacic} is the place to visit.\\

\section{NVEs for the non-Abelian T-dual of AdS$_3\,\times\,$S$^3\times\,$T$^4$}\label{appendixNATD}
Since the particular choice of parameters $c_1=c_4=0$ gives an integrable structure, this should be reflected on the corresponding $r$ and $\chi$ NVEs. Indeed, replacing this particular choice into the NVE for $r$, (\ref{newNVEr}), the latter becomes

\begin{equation}
\ddot{\varrho}\:=\:-(1+\nu^2)\,\varrho
\end{equation}\\
which is the harmonic oscillator, integrable as it should. Replacing also into the NVE for $\chi$, (\ref{newNVEx}), we acquire

\begin{equation}
\ddot{x}\:=\:\left[-4\kappa\left(\kappa+\frac{2E}{R^2}\right)-\frac{48R^4E^2}{(R^4+4E^2\tau^2)^2}-\frac{16R^2E\kappa}{R^4+4E^2\tau^2}\right]\,x\label{xNATDfail}
\end{equation}\\
This equation satisfies the first and second Kovacic's criteria, but yet the algorithm fails to solve it. However, this not yet the correctly informed NVE. That is, since $c_1=c_4=0$ reduce the AdS warp factor to a constant, $f_1=\frac{c_3}{\sqrt{c_5F_0}}=R^2$, then the $t$ equation of motion in \ref{EOM} is solved\footnote{Equivalently, we can find the energy from the worldsheet conjugate momentum as
\begin{equation}
E\:=\:p_0\:=\:\int_0^{2\pi}\dd\sigma\mathcal{P}^\tau_0\:=\:-\frac{2\pi}{4\pi\alpha'}2\,g_{00}\,\dot{t}\:\eqa\: \cosh^2r f_1\,\dot{t}\:\;\;\xrightarrow[t=\tau]{r=0}\:R^2
\end{equation}} for the static gauge\footnote{This is a privilege of the current situation, where $g_{00}\vert_{r=0}=-R^2=const.$ . When $g_{00}(\tau)\vert_{r=0}\neq const.$ , then $t$ behaves as $t=\int\frac{E\,\dd\tau}{g_{00}(\tau)\vert_{r=0}}$ and thus $E$ cannot be specified as a constant and must remain as it is in the equation.} $t=\tau$ and gives $E=R^2$ near $r=0$ (around which we fluctuate). Replacing this into (\ref{xNATDfail}), we get

\begin{equation}
\ddot{x}\:=\:\left(\frac{-48-4\kappa(1+4\tau^2)(6+\kappa+4(2+\kappa)\tau^2)}{(1+4\tau^2)^2}\right)\,x
\end{equation}\\
which is now solved by the algorithm\footnote{We omit the solution since it is of substantial size. The curious reader can put the equation in any algebra software to acquire the solution.}, as it should. Note that the above equation is solved for any choice of gauge $t=\lambda\tau$, $\lambda\in\mathbb{C}$ (and thus every energy $E=\lambda R^2$), as it is appropriate for equivalent physics. Also, notice that we did not really pick a value for the energy $E$ $-$ the energy depends on the observer, i.e. the choice of gauge $-$ the background picked it by itself and we just informed the system about it.\\

This was a typical example of the fact that a failure of Kovacic’s algorithm on a parametrized equation does \textit{not} imply absence of Liouvillian solutions. The algorithm failed to solve (\ref{xNATDfail}), before we correlate its parameters $E,R$ through the physical restrictions of the system. In other words, seeing (\ref{xNATDfail}) purely as a parametrized differential equation, knowing nothing about its physics, we would have to enforce Kovacic's analytic algorithm to find that the choice $E=\lambda R^2$ actually leads to a Liouvillian solution.\\

A special case for the above gauge choice is to set $\lambda=0$, i.e. choose a configuration $t=const.$ . Since the worldsheet theory localizes on target space time $t$, this is an instantonic mode of energy $E=0$. Being one of the legitimate configurations of our string embedding in an integrable space, this instanton has to be integrable as well. Indeed, setting $E=0$ in the NATD NVE (\ref{xNATDfail}) we obtain an harmonic oscillator, integrable as it should.\\

\section{The algorithm for the NVE of $r$}\label{appendixB}
In this appendix we apply the algorithm presented in Appendix \ref{appendixA}, to study the $r$ NVE for both supergravity classes. The main body of the article was reserved for the essential string theory considerations that exclude integrability. Here we just present the explicit calculations that lead to the necessary conditions on which those considerations act.

\subsection{Case 1 for Class I}\label{case1classI}
First in line is the supergravity Class I, with the $r$ NVE (\ref{explicitNVEr}). We begin by writing the partial fraction expansion of $\mathcal{V}_\varrho$ as

\begin{equation}
\mathcal{V}_\varrho\:=\:-\left(\frac{c_5F_0E^2}{c_3^2}+\nu^2\right)+\left(\frac{-c_3^2-4c_1c_4}{4c_3^2}\right)\frac{1}{\tau^2}+\frac{5/16}{(\tau+\frac{c_4}{c_5E})^2}+\frac{5/16}{(\tau+\frac{c_1}{F_0E})^2}+\hdots
\end{equation}\\
where the coefficients $\beta_i$ of the pole terms $1/(\tau-\tau_i)^2$ are used to construct the complex numbers $\alpha_i^\pm=\frac{1}{2}\pm\frac{1}{2}\sqrt{1+4\beta_i}$. In our case these become

\begin{equation}
\alpha_1^\pm\:=\:\frac{1}{2}\pm \sqrt{-\frac{c_1c_4}{c_3^2}}\hspace{1.5cm}\alpha_2^\pm\:=\:\alpha_3^\pm\:=\:\left\lbrace\begin{array}{c}
\frac{5}{4}\\
-\frac{1}{4}
\end{array}\right.\label{ClassICase1ai}
\end{equation}\\
Next, we move to the $\tau\rightarrow\infty$ regime and define a rational function $[\sqrt{\mathcal{V}_\varrho}]_\infty$ which here, since $\mathcal{V}_\varrho^\infty$ is of zeroth order, it has to be just a complex number, i.e. $[\sqrt{\mathcal{V}_\varrho}]_\infty=a$. Then $a$ is found by matching terms between $[\sqrt{\mathcal{V}_\varrho}]_\infty^2$ and $\mathcal{V}_\varrho^\infty$ in (\ref{Vrinfinity}), taking the value $a=i\sqrt{\frac{c_5F_0E^2}{c_3^2}+\nu^2}$. As before, letting $\beta_\infty$ be the coefficient of $1/\tau$ in $\mathcal{V}_\varrho^\infty$, we construct the complex numbers $\alpha_\infty^\pm=\frac{\pm\beta_\infty}{2a}$ which are now valued

\begin{equation}
\alpha_\infty^\pm\:=\:\pm\frac{i\,(c_1c_5+c_4F_0)E}{2\sqrt{c_3^2c_5F_0E^2+c_3^4\,\nu^2}}
\end{equation}\\
Stepping forward, we gather all our findings $\alpha_i^\pm,\alpha_\infty^\pm$ and, letting $s(\cdot)$ be the sign function, we define the numbers $d=\alpha_\infty^{s(\infty)}-\sum_i\alpha_i^{s(i)}$. Considering all the possible sign combinations, these are $2^4=16$ complex numbers. Up to some real constants and signs between their terms, these sixteen numbers are \textit{all} of the form\footnote{We write $\sqrt{-c_1c_4}\rightarrow i\sqrt{c_1c_4}$ for convenience in our following considerations.}

\begin{equation}
d\:\sim\: i\sqrt{\frac{c_1c_4}{c_3^2}}\pm\frac{i\,(c_1c_5+c_4F_0)E}{2\sqrt{c_3^2c_5F_0E^2+c_3^4\,\nu^2}}
\end{equation}\\
Kovacic states that $d$ has to be a \textit{non-negative integer} in order for the algorithm to move on to its next stage. If $d$ is never such a number, then Case 1 cannot give a Liouvillian solution. In other words, the above two terms must be real.\\

Under the string theory considerations on subsection \ref{classIcase1}, we conclude that this can never be the case and, thus, Case 1 cannot hold.\\

\subsection{Case 2 for Class I}\label{case2classI}
In this Case, we begin by considering the same pole coefficients $\beta_i$ that made up the $\alpha_i^\pm$ numbers, (\ref{ClassICase1ai}). But now $\beta_i$'s construct the coordinates $E_i=\lbrace 2+k\sqrt{1+4\beta_i}|k=0,\pm2\rbrace\cap\mathbb{Z}$, which in this case read

\begin{equation}
E_1\:=\:\left\lbrace2-4\sqrt{-\frac{c_1c_4}{c_3^2}}\,,\;2\;,\,2+4\sqrt{-\frac{c_1c_4}{c_3^2}}\right\rbrace\hspace{1.5cm}E_2\:=\:E_3\:=\:\lbrace-1,2,5\rbrace
\end{equation}\\
Under the string theory considerations on subsection \ref{classIcase2}, we conclude that $E_1=\lbrace2\rbrace$.\\

Next, since our potential at infinity, $\mathcal{V}_\varrho^\infty$, is of zeroth order, we also define the coordinate $E_\infty=\lbrace0\rbrace$. Then, in analogy with Case 1, we gather the coordinates $E_\infty$, $E_i$ and define the numbers $d=\frac{1}{2}(e_\infty-\sum_ie_i)$, where $e_i\in E_i$ are the particular coordinates. Again, $d$'s have to be non-integers to be acceptable. Considering all the possible coordinate combinations we calculate $3^2=9$ numbers, of which only one is non-negative, i.e. the one for $e_2=e_3=-1$ ($e_\infty=0$ and $e_1=2$ always) that gives $d=0$.\\

Now, since in this Case we actually obtained a single non-integer $d$, $d=0$, we may move to the next step. That consists of forming the rational function $\theta=\frac{1}{2}\sum_i\frac{e_i}{\tau-\tau_i}$, in which we use the particular $e_i$'s that made up $d=0$, i.e. $e_1=2$, $e_2=e_3=-1$. In our case, $\theta$ is

\begin{equation}
\theta\:=\:\frac{1}{\tau}-\frac{1}{2\left(\tau+\frac{c_4}{c_5E}\right)}-\frac{1}{2\left(\tau+\frac{c_1}{F_0E}\right)}\label{thetaclassIappendix}
\end{equation}\\
Next we search for a monic polynomial $P$ of degree $d$ such that

\begin{equation}
P'''+3\theta P''+(3\theta^2+3\theta'-4\mathcal{V}_\varrho)P'+(\theta''+3\theta\theta'+\theta^3-4\mathcal{V}_\varrho\theta-2\mathcal{V}_\varrho')P\:=\:0
\end{equation}\\
Since $d=0$ is our only heritage from the previous step, that means $P=1$ and the question reduces to whether $\theta''+3\theta\theta'+\theta^3-4\mathcal{V}_\varrho\theta-2\mathcal{V}_\varrho'=0$. Replacing $\theta$, (\ref{thetaclassIappendix}), into the latter necessary condition we find out that it is not satisfied. Therefore, Case 2 also fails to provide a Liouvillian solution.\\

\subsection{Case 1 for Class II}\label{case1classII}
We begin by writing the partial fraction expansion of $\mathcal{V}_\varrho$ as

\begin{equation}
\mathcal{V}_\varrho\:=\:-\left(\frac{(c_5^2-F_0^2)E^2}{c_3^2}+\nu^2\right)+\left(\frac{4(c_1^2-c_4^2)-c_3^2}{4c_3^2}\right)\frac{1}{\tau^2}+\frac{5/16}{(\tau-\frac{c_1-c_4}{(c_5-F_0)E})^2}+\frac{5/16}{(\tau+\frac{c_1+c_4}{(c_5+F_0)E})^2}+\dots
\end{equation}\\
where the coefficients $\beta_i$ of the pole terms $1/(\tau-\tau_i)^2$ are used to construct the complex numbers $\alpha_i^\pm=\frac{1}{2}\pm\frac{1}{2}\sqrt{1+4\beta_i}$. Here, these become

\begin{equation}
\alpha_1^\pm\:=\:\frac{1}{2}\pm \sqrt{\frac{c_1^2-c_4^2}{c_3^2}}\hspace{1.5cm}\alpha_2^\pm\:=\:\alpha_3^\pm\:=\:\left\lbrace\begin{array}{c}
\frac{5}{4}\\
-\frac{1}{4}
\end{array}\right.\label{ClassIICase1ai}
\end{equation}\\
Next, we move to the $\tau\rightarrow\infty$ regime and define the rational function $[\sqrt{\mathcal{V}_\varrho}]_\infty$ which here, since $\mathcal{V}_\varrho^\infty$ is of zeroth order, it has to be just a complex number, i.e. $[\sqrt{\mathcal{V}_\varrho}]_\infty=a$. Then $a$ is found by matching terms between $[\sqrt{\mathcal{V}_\varrho}]_\infty^2$ and $\mathcal{V}_\varrho^\infty$ in (\ref{VrinfinityII}), taking the value $a=i\sqrt{\frac{(c_5^2-F_0^2)E^2}{c_3^2}+\nu^2}$. As before, letting $\beta_\infty$ be the coefficient of $1/\tau$ in $\mathcal{V}_\varrho^\infty$, we construct the complex numbers $\alpha_\infty^\pm=\frac{\pm\beta_\infty}{2a}$ which are now valued

\begin{equation}
\alpha_\infty^\pm\:=\:\pm\frac{i\,(c_4c_5-c_1F_0)E}{\sqrt{c_3^2(c_5^2-F_0^2)E^2+c_3^4\,\nu^2}}
\end{equation}\\
We gather all our findings $\alpha_i^\pm,\alpha_\infty^\pm$ and, letting $s(\cdot)$ be the sign function, we define the numbers $d=\alpha_\infty^{s(\infty)}-\sum_i\alpha_i^{s(i)}$. Considering all the possible sign combinations, these are $2^4=16$ complex numbers. Up to some real constants and signs between their terms, these sixteen numbers are \textit{all} of the form\footnote{We write $\sqrt{c_1^2-c_4^2}\rightarrow i\sqrt{c_4^2-c_1^2}$ for convenience in our following considerations.}

\begin{equation}
d\:\sim\: i\sqrt{\frac{c_4^2-c_1^2}{c_3^2}}\pm\frac{i\,(c_4c_5-c_1F_0)E}{\sqrt{c_3^2(c_5^2-F_0^2)E^2+c_3^4\,\nu^2}}\label{dforCase1rIIapp}
\end{equation}\\
Again, $d$ has to be a non-negative integer for Case 1 to produce a Liouvillian solution, which in turn means that the above two terms must be real.\\

Under the string theory considerations on subsection \ref{classIIcase1}, we conclude that this can never be the case and, thus, Case 1 cannot hold.\\

\subsection{Case 2 for Class II}\label{case2classII}
In Case 2, we begin by considering the same pole coefficients $\beta_i$ that made up the $\alpha_i^\pm$ numbers, (\ref{ClassIICase1ai}). But now $\beta_i$'s construct the coordinates $E_i=\lbrace 2+k\sqrt{1+4\beta_i}|k=0,\pm2\rbrace\cap\mathbb{Z}$, which in this case read

\begin{equation}
E_1\:=\:\left\lbrace2-4\sqrt{\frac{c_1^2-c_4^2}{c_3^2}}\,,\;2\;,\,2+4\sqrt{\frac{c_1^2-c_4^2}{c_3^2}}\right\rbrace\hspace{1.5cm}E_2\:=\:E_3\:=\:\lbrace-1,2,5\rbrace
\end{equation}\\
Under the string theory considerations on subsection \ref{classIIcase2}, we conclude that $E_1=\lbrace2\rbrace$.\\

Since the $E_i$'s are exactly the same with the ones of Class I, we again have a single non-negative integer $d=0$ made out of them, while the rational function $\theta=\frac{1}{2}\sum_i\frac{e_i}{\tau-\tau_i}$ now reads
\begin{equation}
\theta\:=\:\frac{1}{\tau}-\frac{1}{2\left(\tau-\frac{c_1-c_4}{(c_5-F_0)E}\right)}-\frac{1}{2\left(\tau+\frac{c_1+c_4}{(c_5+F_0)E}\right)}\label{thetaCase2ClassII}
\end{equation}\\
The same as the last time, $\theta$ should satisfy $\theta''+3\theta\theta'+\theta^3-4\mathcal{V}_\varrho\theta-2\mathcal{V}_\varrho'=0$. In this class too it does not, therefore Case 2 cannot provide us a Liouvillian solution either, for our $r$ NVE (\ref{explicitNVErII}).\\


\begin{thebibliography}{99}



\bibitem{Beisert:2010jr} 
  N.~Beisert {\it et al.},
  Lett.\ Math.\ Phys.\  {\bf 99}, 3 (2012)
  doi:10.1007/s11005-011-0529-2
  [arXiv:1012.3982 [hep-th]].

\bibitem{Torrielli:2016ufi} 
  A.~Torrielli,
  J.\ Phys.\ A {\bf 49}, no. 32, 323001 (2016)
  doi:10.1088/1751-8113/49/32/323001
  [arXiv:1606.02946 [hep-th]].

\bibitem{Zarembo:2017muf}
  K.~Zarembo,
  arXiv:1712.07725 [hep-th].

\bibitem{Lunin:2005jy} 
  O.~Lunin and J.~M.~Maldacena,
  JHEP {\bf 0505}, 033 (2005)
  doi:10.1088/1126-6708/2005/05/033
  [hep-th/0502086].

\bibitem{Sfetsos:2013wia} 
  K.~Sfetsos,
  Nucl.\ Phys.\ B {\bf 880}, 225 (2014)
  doi:10.1016/j.nuclphysb.2014.01.004
  [arXiv:1312.4560 [hep-th]].
  
\bibitem{Delduc:2014kha} 
  F.~Delduc, M.~Magro and B.~Vicedo,
  JHEP {\bf 1410}, 132 (2014)
  doi:10.1007/JHEP10(2014)132
  [arXiv:1406.6286 [hep-th]].

\bibitem{Borsato:2016pas} 
  R.~Borsato and L.~Wulff,
  Phys.\ Rev.\ Lett.\  {\bf 117}, no. 25, 251602 (2016)
  doi:10.1103/PhysRevLett.117.251602
  [arXiv:1609.09834 [hep-th]].


\bibitem{Kovacic} 
  J. J. Kovacic, ''An algorithm for solving second order linear homogeneous differential equations'', J. Symb. Comp 2 (1986) 3–43


\bibitem{Zayas:2010fs} 
  L.~A.~Pando Zayas and C.~A.~Terrero-Escalante,
  JHEP {\bf 1009}, 094 (2010)
  doi:10.1007/JHEP09(2010)094
  [arXiv:1007.0277 [hep-th]].

\bibitem{Basu:2011fw} 
  P.~Basu and L.~A.~Pando Zayas,
  Phys.\ Rev.\ D {\bf 84}, 046006 (2011)
  doi:10.1103/PhysRevD.84.046006
  [arXiv:1105.2540 [hep-th]].
  
\bibitem{Basu:2012ae} 
  P.~Basu, D.~Das, A.~Ghosh and L.~A.~Pando Zayas,
  JHEP {\bf 1205}, 077 (2012)
  doi:10.1007/JHEP05(2012)077
  [arXiv:1201.5634 [hep-th]].
  
\bibitem{Chervonyi:2013eja} 
  Y.~Chervonyi and O.~Lunin,
  JHEP {\bf 1402}, 061 (2014)
  doi:10.1007/JHEP02(2014)061
  [arXiv:1311.1521 [hep-th]].
  
  
\bibitem{Stepanchuk:2012xi} 
  A.~Stepanchuk and A.~A.~Tseytlin,
  J.\ Phys.\ A {\bf 46}, 125401 (2013)
  doi:10.1088/1751-8113/46/12/125401
  [arXiv:1211.3727 [hep-th]].
  
\bibitem{Giataganas:2013dha} 
  D.~Giataganas, L.~A.~Pando Zayas and K.~Zoubos,
  JHEP {\bf 1401}, 129 (2014)
  doi:10.1007/JHEP01(2014)129
  [arXiv:1311.3241 [hep-th]].
  
\bibitem{Giataganas:2014hma} 
  D.~Giataganas and K.~Sfetsos,
  JHEP {\bf 1406}, 018 (2014)
  doi:10.1007/JHEP06(2014)018
  [arXiv:1403.2703 [hep-th]].
  
\bibitem{Asano:2015eha} 
  Y.~Asano, D.~Kawai and K.~Yoshida,
  JHEP {\bf 1506}, 191 (2015)
  doi:10.1007/JHEP06(2015)191
  [arXiv:1503.04594 [hep-th]].
  
\bibitem{Ishii:2016rlk} 
  T.~Ishii, K.~Murata and K.~Yoshida,
  Phys.\ Rev.\ D {\bf 95}, no. 6, 066019 (2017)
  doi:10.1103/PhysRevD.95.066019
  [arXiv:1610.05833 [hep-th]].
  
\bibitem{Hashimoto:2016wme} 
  K.~Hashimoto, K.~Murata and K.~Yoshida,
  Phys.\ Rev.\ Lett.\  {\bf 117}, no. 23, 231602 (2016)
  doi:10.1103/PhysRevLett.117.231602
  [arXiv:1605.08124 [hep-th]].
  
\bibitem{Giataganas:2017guj} 
  D.~Giataganas and K.~Zoubos,
  JHEP {\bf 1710}, 042 (2017)
  doi:10.1007/JHEP10(2017)042
  [arXiv:1707.04033 [hep-th]].
  
\bibitem{Roychowdhury:2017vdo} 
  D.~Roychowdhury,
  JHEP {\bf 1710}, 056 (2017)
  doi:10.1007/JHEP10(2017)056
  [arXiv:1707.07172 [hep-th]].
  
\bibitem{Banerjee:2018ifm} 
  A.~Banerjee and A.~Bhattacharyya,
  JHEP {\bf 1811}, 124 (2018)
  doi:10.1007/JHEP11(2018)124
  [arXiv:1806.10924 [hep-th]].
  
\bibitem{Roychowdhury:2019olt} 
  D.~Roychowdhury,
  JHEP {\bf 1909}, 002 (2019)
  doi:10.1007/JHEP09(2019)002
  [arXiv:1907.00584 [hep-th]].
  
\bibitem{Akutagawa:2019awh} 
  T.~Akutagawa, K.~Hashimoto, K.~Murata and T.~Ota,
  Phys.\ Rev.\ D {\bf 100}, no. 4, 046009 (2019)
  doi:10.1103/PhysRevD.100.046009
  [arXiv:1903.04718 [hep-th]].
  
\bibitem{Nunez:2018ags} 
  C.~Nunez, J.~M.~Penin, D.~Roychowdhury and J.~Van Gorsel,
  JHEP {\bf 1806}, 078 (2018)
  doi:10.1007/JHEP06(2018)078
  [arXiv:1802.04269 [hep-th]].
  
\bibitem{Nunez:2018qcj} 
  C.~Nunez, D.~Roychowdhury and D.~C.~Thompson,
  JHEP {\bf 1807}, 044 (2018)
  doi:10.1007/JHEP07(2018)044
  [arXiv:1804.08621 [hep-th]].

\bibitem{Filippas:2019puw} 
  K.~Filippas, C.~Nunez and J.~Van Gorsel,
  JHEP {\bf 1906}, 069 (2019)
  doi:10.1007/JHEP06(2019)069
  [arXiv:1901.08598 [hep-th]].

\bibitem{Wulff:2019tzh} 
  L.~Wulff,
  JHEP {\bf 1904}, 133 (2019)
  doi:10.1007/JHEP04(2019)133
  [arXiv:1903.08660 [hep-th]].
  
\bibitem{Wulff:2017lxh} 
  L.~Wulff,
  Phys.\ Rev.\ D {\bf 96}, no. 10, 101901 (2017)
  doi:10.1103/PhysRevD.96.101901
  [arXiv:1708.09673 [hep-th]].
  
\bibitem{Wulff:2017vhv} 
  L.~Wulff,
  JHEP {\bf 1802}, 106 (2018)
  doi:10.1007/JHEP02(2018)106
  [arXiv:1711.00296 [hep-th]].
  
  
\bibitem{Wulff:2017hzy} 
  L.~Wulff,
  J.\ Phys.\ A {\bf 50}, no. 23, 23LT01 (2017)
  doi:10.1088/1751-8121/aa70b5
  [arXiv:1702.08788 [hep-th]].

\bibitem{Giataganas:2019xdj} 
  D.~Giataganas,
  arXiv:1909.02577 [hep-th].





\bibitem{Lozano:2019emq} 
  Y.~Lozano, N.~T.~Macpherson, C.~Nunez and A.~Ramirez,
  arXiv:1908.09851 [hep-th].

\bibitem{Lozano:2019jza} 
  Y.~Lozano, N.~T.~Macpherson, C.~Nunez and A.~Ramirez,
  arXiv:1909.09636 [hep-th].
  
\bibitem{Lozano:2019zvg} 
  Y.~Lozano, N.~T.~Macpherson, C.~Nunez and A.~Ramirez,
  arXiv:1909.10510 [hep-th].
  
\bibitem{Lozano:2019ywa} 
  Y.~Lozano, N.~T.~Macpherson, C.~Nunez and A.~Ramirez,
  arXiv:1909.11669 [hep-th].
  
\bibitem{Speziali:2019uzn} 
  S.~Speziali,
  arXiv:1910.14390 [hep-th].
  
\bibitem{Dibitetto:2017klx} 
  G.~Dibitetto and N.~Petri,
  JHEP {\bf 1801}, 039 (2018)
  doi:10.1007/JHEP01(2018)039
  [arXiv:1707.06154 [hep-th]].
  
\bibitem{Dibitetto:2018iar} 
  G.~Dibitetto and N.~Petri,
  JHEP {\bf 1901}, 193 (2019)
  doi:10.1007/JHEP01(2019)193
  [arXiv:1807.07768 [hep-th]].
  
\bibitem{Hanany:2018hlz} 
  A.~Hanany and T.~Okazaki,
  JHEP {\bf 1903}, 027 (2019)
  doi:10.1007/JHEP03(2019)027
  [arXiv:1811.09117 [hep-th]].
  
\bibitem{Sfetsos:2010uq} 
  K.~Sfetsos and D.~C.~Thompson,
  Nucl.\ Phys.\ B {\bf 846}, 21 (2011)
  doi:10.1016/j.nuclphysb.2010.12.013
  [arXiv:1012.1320 [hep-th]].
  
  
\bibitem{Wulff:2013kga} 
  L.~Wulff,
  JHEP {\bf 1307}, 123 (2013)
  doi:10.1007/JHEP07(2013)123
  [arXiv:1304.6422 [hep-th]].
  
\bibitem{Babichenko:2009dk} 
  A.~Babichenko, B.~Stefanski, Jr. and K.~Zarembo,
  JHEP {\bf 1003}, 058 (2010)
  doi:10.1007/JHEP03(2010)058
  [arXiv:0912.1723 [hep-th]].
  
  
\bibitem{Sfetsos:2010uq} 
  K.~Sfetsos and D.~C.~Thompson,
  Nucl.\ Phys.\ B {\bf 846}, 21 (2011)
  doi:10.1016/j.nuclphysb.2010.12.013
  [arXiv:1012.1320 [hep-th]].

\bibitem{Gaiotto:2009gz} 
  D.~Gaiotto and J.~Maldacena,
  JHEP {\bf 1210}, 189 (2012)
  doi:10.1007/JHEP10(2012)189
  [arXiv:0904.4466 [hep-th]].


\bibitem{Apruzzi:2013yva} 
  F.~Apruzzi, M.~Fazzi, D.~Rosa and A.~Tomasiello,
  JHEP {\bf 1404}, 064 (2014)
  doi:10.1007/JHEP04(2014)064
  [arXiv:1309.2949 [hep-th]].
   
\bibitem{Apruzzi:2015wna} 
  F.~Apruzzi, M.~Fazzi, A.~Passias, A.~Rota and A.~Tomasiello,
  Phys.\ Rev.\ Lett.\  {\bf 115}, no. 6, 061601 (2015)
  doi:10.1103/PhysRevLett.115.061601
  [arXiv:1502.06616 [hep-th]].
  
  

 
\end{thebibliography}
\end{document}